\documentclass[a4paper,11pt]{article}
\usepackage{jheppub}
\usepackage{lineno}
\usepackage{microtype}
\usepackage[T1]{fontenc}
\usepackage{eurosym}
\usepackage{booktabs}
\usepackage{multirow}
\usepackage{graphicx}
\usepackage{amsmath}
\usepackage{colortbl}
\usepackage{amssymb}
\usepackage{todonotes}
\usepackage{wrapfig}
\definecolor{DeepPink}{rgb}{0.796,0.004,0.384}
\definecolor{RoyalBlue}{rgb}{0.02,0.016,0.667}
\definecolor{LavenderBlush}{rgb}{1.00,0.941,0.961}
\usepackage{lastpage}
\usepackage{enumitem}
\usepackage{lipsum}
\usepackage{physics}
\usepackage{longtable}
\usepackage[toc]{multitoc}
\usepackage{comment}
\usepackage{array}

\newcommand{\tauf}{\tau_\mathrm{F}}

\newcommand{\del}[1]{\vphantom{#1}}

\begin{document}
\title{Scale Setting and Strong Coupling Determination in the Gradient Flow Scheme for 2+1 Flavor Lattice QCD}

\author[a]{Rasmus Norman Larsen}
\author[b]{Swagato Mukherjee}
\author[b]{Peter Petreczky}
\author[c]{Hai-Tao Shu}
\author[d,f]{Johannes Heinrich Weber}

\affiliation[a]{Fakult\"at f\"ur Physik, Universit\"at Bielefeld, D-33615 Bielefeld, Germany}
\affiliation[b]{Physics Department, Brookhaven National Laboratory, Upton, New York 11973, USA}
\affiliation[c]{Key Laboratory of Quark \& Lepton Physics (MOE) and Institute of Particle Physics, Central China Normal University, Wuhan 430079, China}
\affiliation[d]{Institut f\"ur Physik \& IRIS Adlershof, Humboldt-Universit\"at zu Berlin, D-12489 Berlin, Germany}
\affiliation[f]{Institut f\"ur Kernphysik, Technische Universit\"at Darmstadt, 
Schlossgartenstra\ss e 2, D-64289 Darmstadt, Germany}

\emailAdd{rlarsen@physik.uni-bielefeld.de, swagato@bnl.gov,\\
petreczk@bnl.gov, hai-tao.shu@ccnu.edu.cn, dr.rer.nat.weber@gmail.com}

\abstract{
We report on the determination of the gradient flow scales in  $N_f=2+1$ QCD using highly improved staggered quark (HISQ) ensembles generated by the HotQCD Collaboration for bare gauge couplings ranging from $\beta = 6.423$ to $8.400$. Using bottomonium splittings, kaon decay constant, the decay 
constant of unmixed $\eta_s$ meson and the $\phi$ meson mass we obtained the values of the gradient
flow scales in physical units, $\sqrt{t_0} = 0.14446 (76)$~fm and $w_0 = 0.17429 (83)$~fm. 
Using the same physical inputs we revisit the determination of the potential $r_1$ scale 
and find $r_1 = 0.3118(26)$~fm. As a byproduct of our study we obtain the running of the gauge coupling
in the gradient flow scheme. We find that within the uncertainties the running 
of the gradient flow coupling obtained on the lattice is compatible with the perturbative results
up to flow radius $\sqrt{8 \tau_F}=0.15$ fm.
}
\date{\today}
\maketitle

\section{Introduction}

Scale setting is an essential step in the lattice calculations that converts the lattice computed dimensionless quantity to dimensionful quantity in physical units. It affects the accuracy of lattice calculations in such a fundamental way that it must be handled as precisely as possible. There are different strategies to this end, which roughly divide into two categories. One is to use the physical scales that are experimentally accessible, the other is to use the theory scales. The former mainly includes the masses of baryons (e.g. $\Omega$) \cite{RBC:2014ntl,Miller:2020evg,BMW:2012hcm,RBC:2010qam,PACS-CS:2008bkb} and the pion/kaon decay constant $f_\pi$/$f_K$, see, e.g. \cite{RBC:2014ntl,RBC:2010qam,PACS-CS:2008bkb,Davies:2009tsa}. The latter mainly includes the (static-quark) potential scales $r_0$ and $r_1$ \cite{Sommer:1993ce,Bernard:2000gd,Francis:2015lha,HotQCD:2014kol,Bazavov:2011nk,Brambilla:2022het}, the decay constant $F_{p4s}$ of fictitious pseudo-scalar meson with mass $0.4$ times the strange quark mass \cite{MILC:2012znn,MILC:2015tqx}, and the gradient flow scales $t_0$ and $w_0$ \cite{Miller:2020evg,BMW:2012hcm,ExtendedTwistedMass:2021qui,Borsanyi:2020mff,MILC:2015tqx,Bruno:2016plf,BMW:2012hcm,HotQCD:2014kol,RBC:2014ntl,Dowdall:2013rya,Bornyakov:2015eaa} which we will pursue in this work. 
Calculating theory scales on the lattice is often much easier than calculating 
physical quantities and therefore, these are useful to determine the lattice
spacing in practical lattice QCD calculations. One prominent example is calculation at nonzero temperature, which often requires simulations at many lattice spacings. Because the theory scales themselves have no direct physical meaning, their ratios to physical observables must first be determined at each lattice spacing. These ratios are then extrapolated to the continuum limit, from which the theory scale in physical units is obtained. In principle, any dimensionful quantity can be used to ``define" a theory scale, as long as it is finite in the continuum limit. But in practice it is only worth to consider those that are cheap and straightforward to compute on the lattice. That is, these quantities must be precise enough that with a small computational cost one can acquire good control on the statistical and systematic uncertainties.

Among them are the commonly used potential scales $r_0$ and $r_1$ proposed  decades ago \cite{Sommer:1993ce,Bernard:2000gd}. They are defined by $r^2\mathrm{d}V(r)/\mathrm{d}r=1.65$ or $1$ for $r=r_0$ or $r_1$, respectively. Usually the potential $V(r)$ itself is precise on the lattice, but when the lattice is very fine, the statistical uncertainties of the Wilson loops at large temporal distances (in lattice units) become significant, which weakens the control of the excited-state contamination. $F_{p4s}$, on the other hand, has a relatively strong dependence on the valence quark mass. To obtain $F_{p4s}$ one needs to perform
fits to meson correlation function at asymptotically large separations, which again could be challenging for small lattice spacing.

Gradient flow \cite{Narayanan:2006rf,Luscher:2009eq} is a seminal framework that has applications in many fields, such as renormalization, defining operators, noise reduction, scale setting \textit{etc}. The flow scales are introduced based on a crucial property of gradient flow that 
it renormalizes composite operators at positive finite flow time
\cite{Luscher:2011bx}.  For gradient flow scales the relevant composite operator is the vacuum gauge action density $E$. The flow scales have been a topic of intense study for the last decade for their stability and high precision. This is due to the fact that the action density is very precise on the lattice and no fits of correlation functions at large separations are needed.

The gradient flow scales $t_0$ and $w_0$ are the most commonly used at present.
Therefore, it is important to obtain precise and reliable results for them.
Cross comparison of different lattice QCD calculations of $t_0$ and $w_0$ can
help to reach this goal. The number of independent determination of these scales
in 2+1+1 flavor QCD is limited because several determinations use the same set
of gauge configurations. There is some tension between the determination of
the $t_0$ and $w_0$ scale obtained using rooted staggered fermions \cite{Dowdall:2013rya,MILC:2015tqx,Miller:2020evg,Bazavov:2025mao}
and twisted mass fermions \cite{ExtendedTwistedMass:2021qui}. For 2+1 flavor QCD the Flavor Lattice Averaging Group (FLAG) quotes
a $t_0$ value with very small error in their 2024 report. However, there is a significant scattering 
in different $t_0$ determination that enter the FLAG 2024 report \cite{FlavourLatticeAveragingGroupFLAG:2024oxs}.
Calculating both $w_0$ and $t_0$ as well as their ratio
could provide an important consistency check. However, many studies present only 
results on either $t_0$ or $w_0$. Furthermore, it may be also interesting to
study the ratio of gradient flow scales and the potential scales. Unfortunately,
very few studies are available for such ratios. 
An interesting question is whether the gradient flow scales are different in 2+1
flavor and 2+1+1 flavor QCD. The FLAG report tends to suggest that this is the case \cite{FlavourLatticeAveragingGroupFLAG:2024oxs}, but
the calculation by ETMC in 2+1+1 flavor QCD with twisted mass Wilson
fermions arrives at the opposite conclusion \cite{ExtendedTwistedMass:2021qui}.

The goal of this work is to determine the gradient flow scales $t_0$ and $w_0$
as well as their ratio to the $r_1$ scale in 2+1 flavor QCD using the HISQ action
and the gauge configurations generated at many different lattice spacings \cite{HotQCD:2014kol,Bazavov:2017dsy,Altenkort:2023oms}.
One of the motivations of this study is scale setting for calculations at high temperature
in 2+1 flavor QCD with HISQ action \cite{Bazavov:2013uja,Ding:2015fca,Bazavov:2016uvm,Bazavov:2018wmo,Altenkort:2023eav,Bazavov:2023dci,Ding:2025fvo,HotQCD:2025fbd}.
In these calculations the scale was almost exclusively fixed by $r_1$. It would be good to have an alternative way for 
determining the scale for these calculations.

There are not many independent determinations of the $r_1$ scale as many 
of the existing determinations use the same set of rooted staggered gauge configurations.
There is tension between some of the determinations, including the recent TUMQCD
determination \cite{Brambilla:2022het}. 
On the other hand the very recent analysis with clover fermion formulation \cite{Asmussen:2024hfw} finds
 that $r_1$ is in perfect agreement with the FLAG average.
Therefore, we also revisit the determination
of $r_1$ scale in this work and compare it with the very recent result from Ref. \cite{Asmussen:2024hfw}.

The gradient flow also provides a convenient definition of finite-volume running couplings~\cite{Luscher:2010iy,Fodor:2012td,Fritzsch:2013je}, which have been used for precise nonperturbative determinations of the QCD $\Lambda$ parameter in three-flavor QCD~\cite{DallaBrida:2016kgh,Bruno:2017gxd}, as well as in pure SU(3) Yang--Mills theory using the twisted gradient-flow coupling~\cite{Ishikawa:2017xam}. Recent applications of this approach can be found in, e.g. \cite{Hasenfratz:2023bok,Wong:2023jvr}. Therefore, in this work we will also calculate the gradient flow coupling 
in the continuum limit at small flow time.

This paper is organized as follows. We start with a short summary of the theoretical foundations, including the gradient flow equations, the definitions of the gradient flow scales and the Allton-type ansatz that relates the flow scales to $\beta$ in Sec. \ref{theory}. In Sec. \ref{latticesetup} we provide the lattice setup used in this study. In Sec. \ref{sec:scales} we present our results for the flow scales, the comparisons of the ratios of scales with the literature and the relation that determines the lattice spacing $a$ given $\beta$ via flow scales. 
In Sec. \ref{results} we present the results for the flow scales and potential scales in physical units. 
In Sec. \ref{LambdaMSbar} we discuss
the QCD running coupling constant in the gradient flow scheme in the
continuum limit and comparison with the perturbative running. We summarize our findings in Sec. \ref{summary}.

\section{Formalism}
\label{theory}
Yang-Mills gradient flow evolves the gauge fields along the gradient of the gauge action according to the following diffusion equations \cite{Narayanan:2006rf,Luscher:2009eq}
\begin{align}
\label{eq:flowdef}
    B_\mu(x,\tauf=0) & = A_\mu(x), \\ 
    \partial_{\tauf} B_\mu(x,\tauf) & = D_\nu G_{\nu\mu} ,
\end{align}
where $A_\mu(x)$ is the original gauge field and $B_\mu(x,\tauf)$ denotes the gauge field flowed to flow time $\tauf$. $D_\nu$ and $G_{\nu\mu}(x)$ are the flowed covariant derivative and field strength constructed using $B_\mu(x,\tauf)$. On the lattice the flow equations take the form
\begin{align}
\label{eq:flowlat}
    V(x,\mu)|_{\tauf=0}&=U(x,\mu), \\
    \frac{\mathrm{d}}{\mathrm{d}\tauf}V(x,\mu)&=-g_0^2 \{\partial_{x,\mu}S_G(V) \}V(x,\mu),
\end{align}
where $U(x,\mu)$ is the original gauge link and $V(x,\mu)$ is its flowed version. $g_0$ is the bare coupling and $S_G$ can take different forms upon how the gauge action is discretized. In this work we use the \textit{Zeuthen flow}~\cite{Ramos:2015baa} that adopts a Symanzik improved L$\ddot{\mathrm{u}}$scher–Weisz gauge action \cite{Luscher:1984xn, Luscher:1985zq} eliminating tree-level $\mathcal{O}(a^2)$ discretization effects  \cite{Ramos:2015baa}.

The leading-order solutions of the flow equations in Eq.~(\ref{eq:flowdef}) can be expressed in terms of a $\tauf$-dependent transformation of the gauge field $B_\mu(x,\tauf)$~\cite{Luscher:2010iy}:
\begin{align}
    \label{eq:flow_solution}
    \begin{split}
        &B_{\mu}(x,\tauf)=\int \mathrm{d}^4 y\, K_{\tauf}(x-y)\, A_{\mu}(y),\\
        &K_{\tauf}(z)=\int \frac{\mathrm{d}^4 p}{(2\pi)^4}\, e^{ipz}\, e^{-\tauf p^2}
        =\frac{e^{-z^{2}/(4\tauf)}}{(4\pi \tauf)^{2}},
    \end{split}
\end{align}
where $K_{\tauf}$ is a local Gaussian smearing kernel with radius $\sqrt{8\tauf}$~\cite{Luscher:2011bx}.
The smearing removes the short-distance fluctuations of the gauge fields and has been widely used as a noise reduction technique, see e.g. \cite{Altenkort:2020fgs,Altenkort:2020axj,Brambilla:2022xbd,Altenkort:2022yhb}. In this work we use it to define non-perturbative reference scales, which involve a very simple gauge invariant operator, the action density, that reads
\begin{align}
    \label{eq:action-density}
       E(\tauf,x)=-\frac{1}{2}\mathrm{Tr}\{G_{\mu\nu}(\tauf,x)G_{\mu\nu}(\tauf,x)\}.
\end{align}
In practical simulations an average over all space-time position $x$ is taken. 

The gradient flow scales $t_i$ and $w_i$ can be defined by imposing the conditions \cite{Luscher:2010iy,BMW:2012hcm}
\begin{equation}
    \begin{split}
        \tauf^2\langle E(\tauf)\rangle\Big{|}_{\tauf=t_i} &= c_i,\\
     \tauf\frac{\mathrm{d}}{\mathrm{d}\tauf}\left\{\tauf^2\langle E(\tauf)\rangle\right\}\Big{|}_{\tauf=w_i^2} &= c_i, i=0,1,2.
    \end{split}
    \label{eq:fixpoint}
\end{equation}
The commonly used choice $c_0=0.3$ defines the scales $t_0$ and $w_0$ that are often used in the lattice calculations.
The alternative choice $c_1=0.7$ has been used by Wuppertal-Budapest Collaboration to define $w_1$ scale \cite{Borsanyi:2023wno}.
In this work we also consider the choice $c_2=0.2$ that defines scales $t_2$ and $w_2$.  Note that in lattice calculations all quantities are in lattice units, thus what one can obtain from these two conditions are actually two dimensionless quantities $\sqrt{t_i}/a$ and $w_i/a$. 
In this study, we consider two different discretizations for the action density $E$, in terms of two different discretization for the field strength tensor $G_{\mu\nu}$. One is to use the clover-type discretization
\begin{align}
    \label{eq:Fmunu}
    \hat{G}^{\mathrm{clo}}_{\mu\nu}(n) = -\frac{i}{8}\Big{(}Q_{\mu\nu}(n)-Q_{\nu\mu}(n)\Big{)},
\end{align}
where $Q_{\mu\nu}(n)$ is sum of four square plaquettes, see, e.g. Ref. \cite{Gattringer:2010zz}. The operator is hatted to indicate that it is in lattice units. The other is based on an $a^2$-improved discretization that employs a mixture of square and $1\times 2$ rectangle plaquettes 
\begin{align}
    \label{eq:impF}
    \hat{G}^{\mathrm{imp}}_{\mu\nu}(n) = \frac{5}{3}C^{(1,1)}_{\mu\nu}(n)-\frac{1}{3}C^{(1,2)}_{\mu\nu}(n),
\end{align}
see e.g. Ref. \cite{Bilson-Thompson:2002xlt} for more details. 
The use of two different discretization schemes for the action density provides important consistency
checks in the determination of the gradient flow scales as will be  explained below.

\begin{figure*}[tbh]
\centerline{
\includegraphics[width=0.5\textwidth]{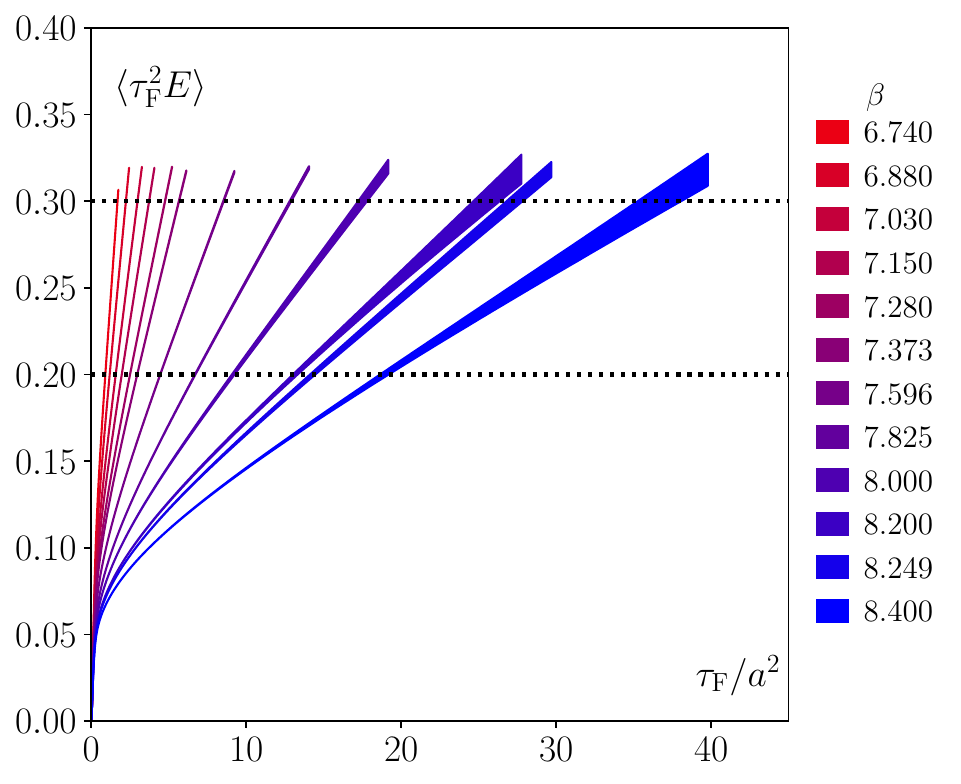}
\includegraphics[width=0.5\textwidth]{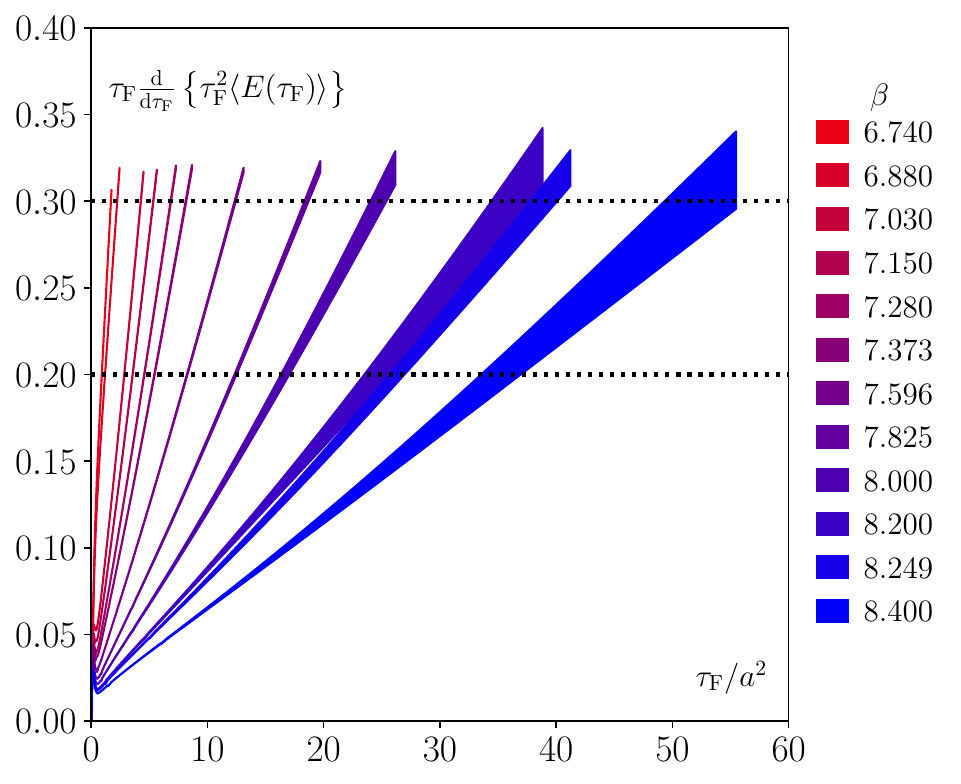}
}
\caption{Numerical results for the action density, $E$ (left) and the derivative of $\tauf^2 E$(right) as function of the flow time obtained using the improved discretization.
}
\label{fig:t2E}
\end{figure*}

Once the flow scales are obtained, they are fitted to the following Allton-type Ansatz \cite{Allton:1996dn} as a function of $\beta$
\begin{equation}
\begin{split}
& \frac{\sqrt{t_i}}{a}=\frac{1+d_{2i}(10/\beta)f^2(\beta)}{c_{0i}f(\beta)+c_{2i}(10/\beta)f^3(\beta)},\\
& \frac{w_i}{a}=\frac{1+d'_{2i}(10/\beta)f^2(\beta)}{c'_{0i}f(\beta)+c'_{2i}(10/\beta)f^3(\beta)},\\
&f(\beta)=\left(\frac{10b_0}{\beta}\right)^{-b_1/(2b_0^2)}\exp(-\beta/(20b_0)),
\end{split}
\label{eq:fit-allton}
\end{equation}
where $\beta=10/g_0^2$, $b_0=9/(4\pi)^2$ and $b_1=1/(4\pi^4)$ for $N_f=3$. 
($d_{2i}$, $c_{0i}$, $c_{2i}$) and ($d'_{2i}$, $c'_{0i}$, $c'_{2i}$)
are fit parameters. Using $t_0$ and $w_0$ in physical units obtained through 
, e.g. bottomonium mass splitting, the lattice spacing can be predicted for a new $\beta$.

\section{Gradient flow scales in 2+1 flavor lattice QCD}
\subsection{Lattice setup}
\label{latticesetup}
\begin{table}[htb]
\centering
\begin{tabular}{cccccc}
\hline \hline
$\beta$ & $a m_s$ & $a m_l$ & $N_{\sigma}$ & $N_{\tau}$ & \# conf. \\
\hline
6.423	& 0.0670 &	0.00335 & 32 & 32 & 500 \\
6.515	& 0.0604 &	0.00302 & 32 & 32 & 500 \\
6.550	& 0.0582 &	0.00291 & 32 & 32 & 700 \\
6.575	& 0.0564 &	0.00282 & 32 & 32 & 500 \\
6.664	& 0.0514 &	0.00257 & 32 & 32 & 700 \\
6.740	& 0.0476 &	0.00238 & 48 & 48 & 700 \\
6.880	& 0.0412 &	0.00206 & 48 & 48 & 627 \\
7.030	& 0.0356 &	0.00178 & 48 & 48 & 900 \\
7.150	& 0.0320 &	0.00160 & 48 & 64 & 395 \\
7.280	& 0.0280 &	0.00142 & 48 & 64 & 398 \\
7.373	& 0.0250 &	0.00125 & 48 & 64 & 554 \\
7.596	& 0.0202 &	0.00101 & 64 & 64 & 577 \\
7.825	& 0.0164 &	0.0082 & 64 & 64 & 471 \\
\hline
7.030	& 0.0356 &  0.00712 & 48 & 48 & 280 \\
7.825	& 0.01542 &	0.003084 & 64 & 64 &  160 \\
8.000	& 0.01299 &	0.002598 & 64 & 64 & 1004 \\
8.200 & 0.01071 & 0.002142 & 64 & 64 & 961 \\
8.249 & 0.01011 & 0.002022 & 64 & 64 & 2241 \\
8.400 & 0.00887 & 0.001774 & 64 & 64 & 2372 \\
\hline \hline
\end{tabular}
\caption{Parameters of the lattice ensembles used in this work, including the inverse gauge coupling $\beta=10/g_0^2$, the bare quark masses in lattice units, the spatial and temporal lattice extent $N_\sigma$ and $N_\tau$, and the corresponding number of configurations.}
\label{tab:param-zeroT}
\end{table}

The calculations are carried out on the ensembles generated by the HotQCD Collaboration in Refs. \cite{HotQCD:2014kol,Bazavov:2017dsy,Altenkort:2023oms} using 2+1 flavor HISQ action \cite{Follana:2006rc} and tree-level improved L$\ddot{\mathrm{u}}$scher-Weisz gauge action \cite{Luscher:1984xn,Luscher:1985zq}. Seventeen different 
values of the lattice gauge coupling $\beta=10/g_0^2$ are employed in this study, ranging from 6.423 to 8.400.
The calculations have been performed for the strange quark mass, $m_s$ close to its physical value, while the degenerate light quark masses were fixed
to be either $m_s/m_l=20$ or $m_s/m_l=5$.
For the largest $\beta$ value additional gauge configurations have been generated to improve the determinations of the gradient flow scales.
The details are summarized in Tab.~\ref{tab:param-zeroT}.  In the first block of Tab.~\ref{tab:param-zeroT} ($6.423\leq \beta \leq 7.825$), the light quark masses are fixed as a fraction of the strange quark mass $m_s/m_l=20$, slightly above the physical ratio $m_s/m_l=27.3$. This corresponds to a pion mass of about 160 MeV in the continuum limit. 
The second block 
corresponds to the light quark mass of $m_s/m_l=5$ and
$7.030 \leq \beta \leq 8.400$. This light quark mass is equivalent to the pion mass of about $320$ MeV in
the continuum limit.
These gauge configurations have been used as the zero temperature reference for the calculations at high temperature \cite{Bazavov:2017dsy,Bazavov:2023dci,Altenkort:2023oms} as well as for 
the determination of the heavy quark masses and strong coupling constant through the moments of quarkonium correlators \cite{Petreczky:2019ozv,Petreczky:2020tky}.
As discussed in Ref. \cite{Bazavov:2017dsy} the generated gauge configurations for $\beta = 8.0$, 8.2 and 8.4 are at fixed topology for a given Monte-Carlo
stream. However, we have streams that belong to different values of topological charge \cite{Bazavov:2017dsy} allowing us to estimate 
the effects of frozen topology for the quantities of interest as discussed below.

\begin{table*}[tbh]
\centering
\renewcommand{\arraystretch}{2.0}
\begin{tabular}{c|c|c|c|c}
\hline \hline
 & $c_{00}$ & $c_{20}/10^{5}$ &  $d_{20}/10^{3}$ &  $\chi^2$/d.o.f.  \\ \hline 
$\sqrt{t_0}/a$  
 & 93.2767(0.0747) & 5.9633(0.1149) & 3.9169(0.0975) & 9.8  \\ \hline
& $c'_{00}$ & $c'_{20}/10^{5}$ &  $d'_{20}/10^{3}$ &  $\chi^2$/d.o.f.  \\ \hline 
$w_0/a$ 
 & 78.0681(0.1204) & 2.9399(0.1236) & 1.5429(0.1191) & 22.2  \\ \hline
\hline
 & $c_{00}$ & $c_{20}/10^{6}$ & $d_{20}/10^{4}$ & $\chi^2$/d.o.f. \\ \hline 
$\sqrt{t_0}/a$  
 &  88.2315(1.6795) & 4.4900(1.7799) & 4.0278(1.6889) & 0.5 \\ \hline
&  $c'_{00}$ & $c'_{20}/10^{6}$ &  $d'_{20}/10^{4}$ & $\chi^2$/d.o.f. \\ \hline 
$w_0/a$ 
 & 74.6393(2.8458) & 2.4417(2.3542) & 2.4608(2.6035) & 0.5 \\ \hline
\hline
\end{tabular}
\caption{Fit parameters of the gradient flow scales obtained from the improved discretization to the Allton Ansatz. Here we rescale the fit parameters to make the values more readable. We only consider the improved discretization. The top block is for the $m_s/m_l=20$ ensembles while the bottom block is for the $m_s/m_l=5$ ensembles.}
\label{tab:fit-scale-setting}
\end{table*}

\begin{figure*}[tbh]
\centerline{
\includegraphics[width=0.5\textwidth]{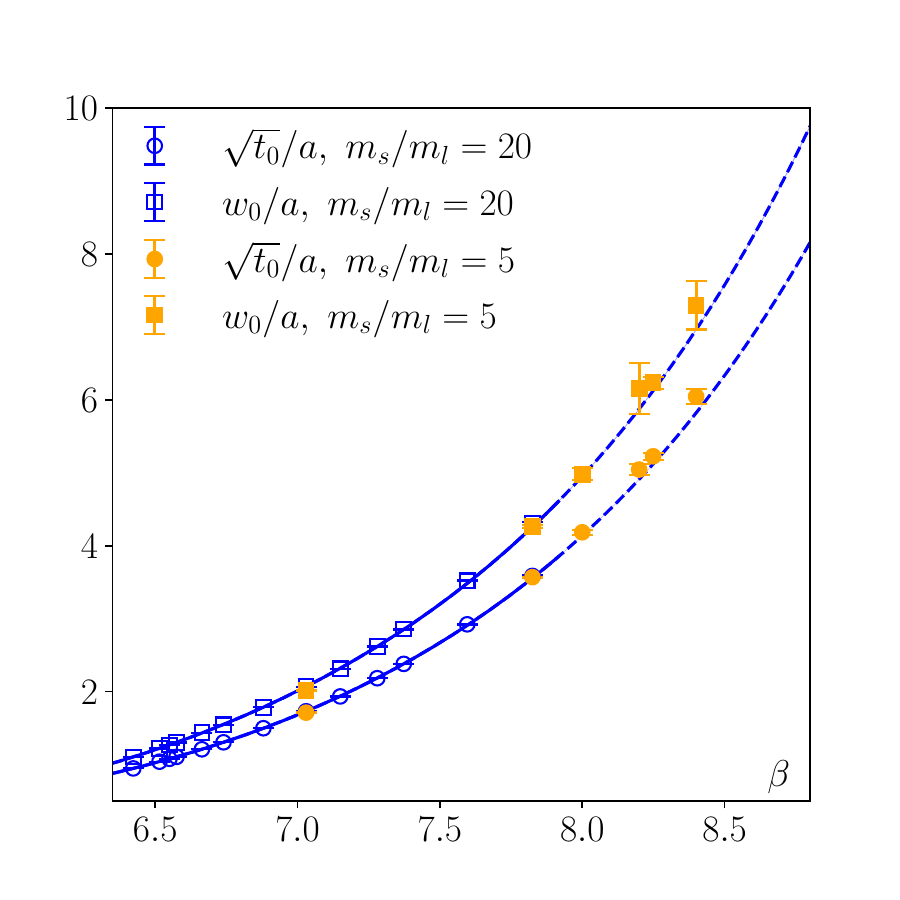}
}
\caption{The gradient flow scales $\sqrt{t_0}$ and $w_0$ obtained from the improved discretization as functions of $\beta$. The open points in blue are for the $m_s/m_l=20$ lattices while the filled points in orange are for the $m_s/m_l=5$ lattices. The solid lines (means) and bands (errors) represent the fits to Allton Ansatz of the $m_s/m_l=20$ lattices ($6.423\leq \beta\leq 7.825$). 
The dashed lines represent the extrapolation of the fits to the $m_s/m_l=5$ lattices, to test the robustness of the fit by comparing its extrapolated predictions with data not included in the fitting procedure.
}
\label{fig:flow-scales-setting}
\end{figure*}

\subsection{Determination of the gradient flow scales}
\label{sec:scales}

\begin{table*}[tbh]
\centering
\setlength{\tabcolsep}{5pt}
\tiny
\begin{tabular}{c|c|c|c|c}
\hline  \hline  
& \multicolumn{1}{c|}{$\sqrt{t_0}/w_0$} & \multicolumn{1}{c|}{$\sqrt{t_2}/w_2$} & \multicolumn{1}{|c|}{$\sqrt{t_0}/\sqrt{t_2}$} & \multicolumn{1}{c}{$w_0/w_2$}  \\ \hline 
  &  $a^2$   &  $a^2$   &  $a^2$   &  $a^2$ \\ \hline 
8 largest $\beta$ & 0.8390(7), 3.0 & 0.7562(7), 22.0 & 1.3580(5), 42.0 & 1.2302(9), 2.8 \\ 
7 largest $\beta$ & 0.8351(14), 1.6 & 0.7478(10), 6.7 & 1.3657(7), 15.6 & 1.2268(15), 1.5 \\ 
6 largest $\beta$ & 0.8337(19), 1.3 & 0.7426(15), 2.6 & 1.3717(10), 4.9 & 1.2256(20), 1.4 \\ 
5 largest $\beta$ & \bf{0.8307(28), 1.0} & \bf{0.7382(20), 1.1} & 1.3762(14), 1.6 & 1.2244(28), 1.5 \\ 
4 largest $\beta$ & 0.8318(44), 1.0 & 0.7357(32), 0.8 & \bf{1.3799(22), 0.5} & 1.2202(40), 1.1 \\ 
3 largest $\beta$ & 0.8331(49), 0.9 & 0.7369(38), 0.5 & 1.3792(27), 0.3 & \bf{1.2194(47), 1.0} \\
\hline 
\end{tabular}
\begin{tabular}{c|c|c|c|c|c|c}
\hline 
& \multicolumn{3}{c|}{$\sqrt{t_0}/w_0$} & \multicolumn{3}{c}{$\sqrt{t_2}/w_2$} \\ \hline
  &  $a^2$  &  $\alpha_b a^2$   & $\alpha_b a^2+a^4$  &  $a^2$  &   $\alpha_b a^2$  & $\alpha_b a^2+a^4$ \\ \hline 
8 largest $\beta$ & 0.8275(7), 1.2 & 0.8289(7), 1.1 & 0.8298(22), 1.1 & 0.7302(7), 1.3 &  0.7323(6), 1.7 & 0.7297(15), 1.3 \\
7 largest $\beta$ & \bf{0.8285(12), 1.0}  & \bf{0.8294(11), 1.0} & 0.8287(33), 1.0 & 0.7297(9), 1.3 & 0.7314(8), 1.5 & 0.7271(24), 1.2 \\
6 largest $\beta$ & 0.8286(20), 1.0 &  0.8293(19), 1.1 & 0.8253(57), 1.1 & 0.7285(13), 1.3 &  0.7299(13), 1.3 & 0.7262(39), 1.4 \\ 
5 largest $\beta$ & 0.8262(32), 0.9 & 0.8269(30), 0.9 & 0.8313(86), 0.9 & \bf{0.7264(19), 1.2} & \bf{0.7277(18), 1.1} & 0.7323(55), 1.1 \\ 
4 largest $\beta$ & 0.8276(47), 1.0 & 0.8280(43), 1.0 & 0.8384(151), 0.7 & 0.7274(37), 1.4 & 0.7282(34), 1.3 & 0.7418(85), 0.2 \\ 
3 largest $\beta$ & 0.8291(64), 1.1 & 0.8293(59), 1.1 & 0.8649(268), 0 & 0.7311(40), 0.5 &  0.7316(37), 0.4 & 0.7453(178), 0 \\ 
\hline 
\end{tabular}
\begin{tabular}{c|c|c|c|c|c|c}
\hline 
& \multicolumn{3}{c|}{$\sqrt{t_0}/\sqrt{t_2}$} & \multicolumn{3}{c}{$w_0/w_2$} \\ \hline 
  &  $a^2$  &   $\alpha_b a^2$  & $\alpha_b a^2+a^4$  &  $a^2$  &  $\alpha_b a^2$ & $\alpha_b a^2+a^4$ \\ \hline 
8 largest $\beta$ & 1.3869(5), 2.2 &  1.3850(5), 2.9 & 1.3880(12), 2.0  & 1.2249(9), 1.1 &  1.2248(8), 1.1 & 1.2228(25), 1.0 \\ 
7 largest $\beta$ & 1.3874(7), 2.3 &  1.3859(7), 2.7 & 1.3904(20), 1.7  & \bf{1.2239(15), 1.0}  & \bf{1.2240(13), 1.0} & 1.2231(37), 1.0 \\ 
6 largest $\beta$ & 1.3888(11), 1.9 &  1.3876(11), 2.1 & 1.3918(31), 1.9  & 1.2235(22), 1.0 &  1.2236(20), 1.0 & 1.2234(59), 1.1  \\ 
5 largest $\beta$ & 1.3906(17), 1.5 & \bf{1.3894(15), 1.5} & 1.3879(43), 1.6  & 1.2241(33), 1.0 & 1.2242(30), 1.0 & 1.2189(93), 1.0  \\ 
4 largest $\beta$ & 1.3907(26), 1.8 & 1.3899(23), 1.7 & 1.3792(65), 0.3  & 1.2220(46), 0.9 &  1.2222(41), 1.0 & 1.2210(161), 1.2  \\ 
3 largest $\beta$ & \bf{1.3880(32), 0.6} & 1.3875(30), 0.5 & 1.3789(119), 0 & 1.2236(69), 1.1  & 1.2237(62), 1.1 & 1.1881(271), 0 \\ 
\hline \hline  
\end{tabular}
\caption{Continuum-extrapolated ratios of the flow scales from different fit strategies using data from the clover (top) and improved discretization (middle and bottom). The numbers are in the form \{Mean(Error), $\chi^2$/d.o.f.\}. }
\label{tab:fit-param-flow-ratio}
\end{table*}

\begin{figure*}[tbh]
\includegraphics[width=0.5\textwidth]{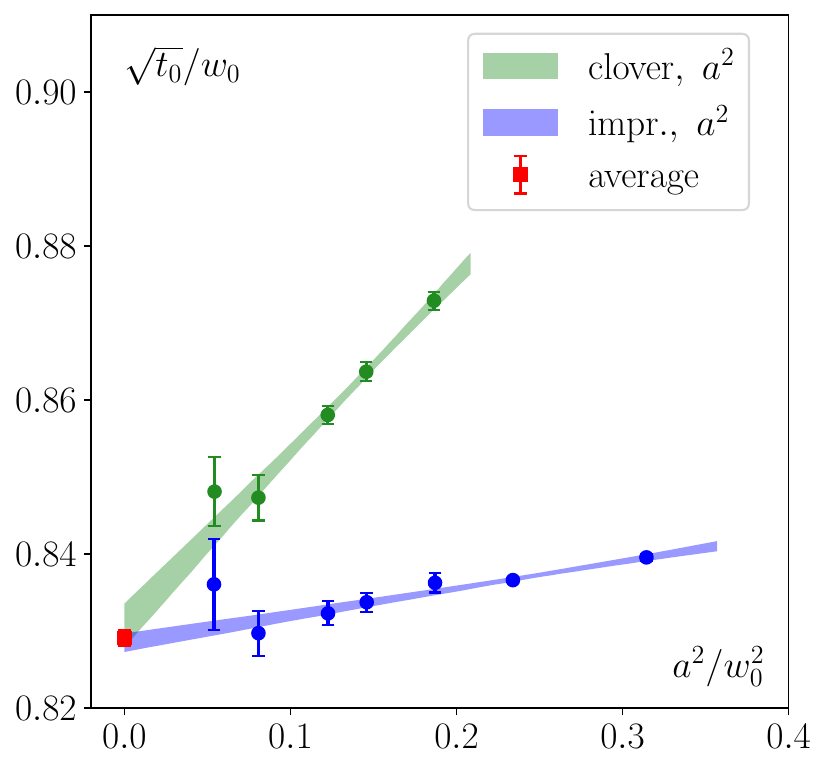}
\includegraphics[width=0.5\textwidth]{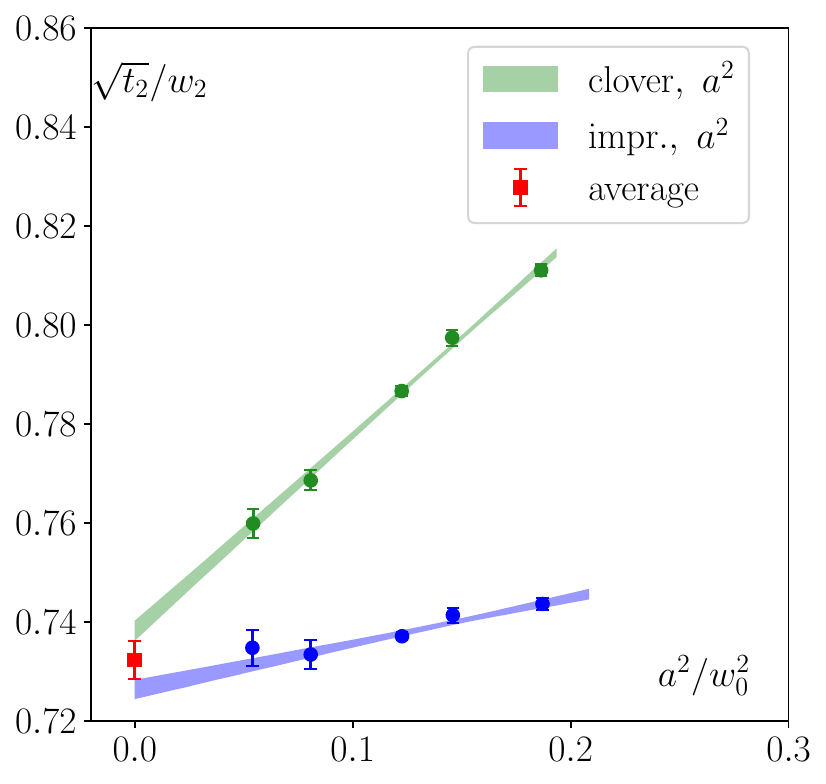}
\includegraphics[width=0.5\textwidth]{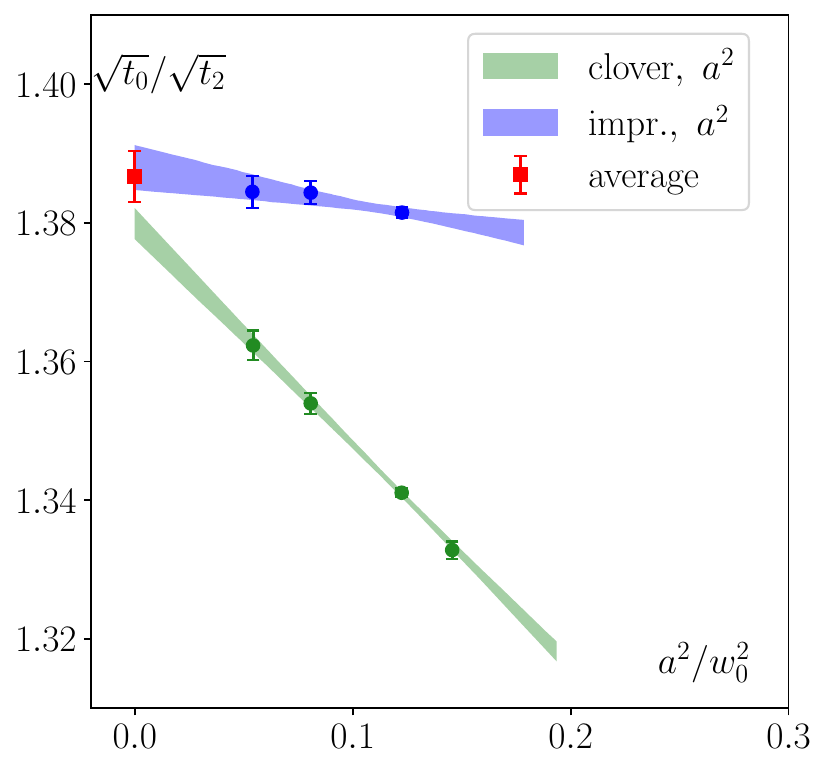}
\includegraphics[width=0.5\textwidth]{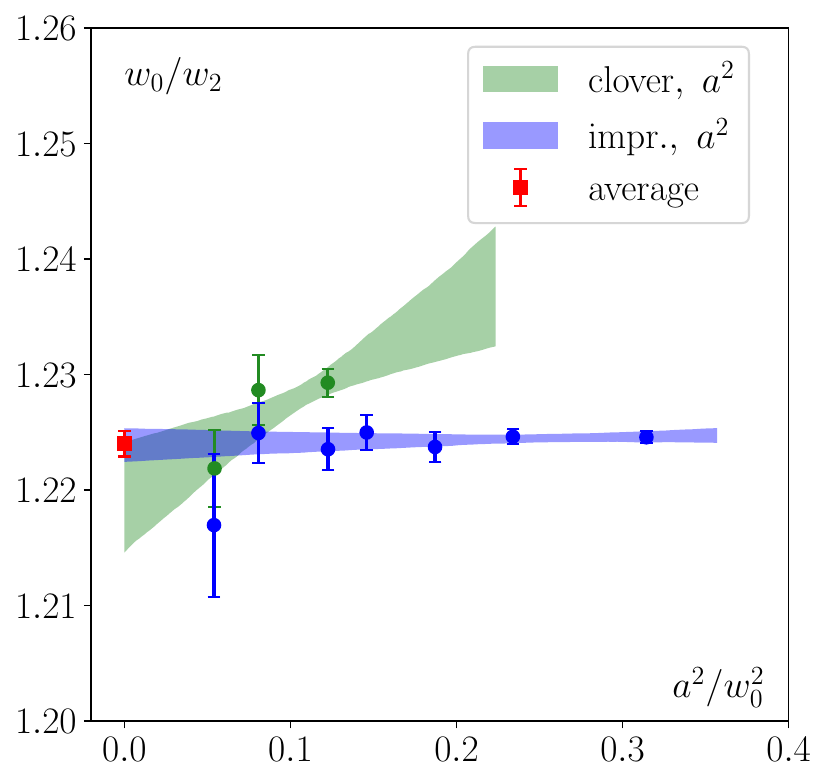}
\caption{
The ratio of different gradient flow scale as function of the lattice spacing
together with the continuum extrapolations. The flow scales are measured using both the clover discretizations and the improved discretization (labelled as ``clover" and ``impr.", respectively, see Eq.~(\ref{eq:Fmunu})-Eq.~(\ref{eq:impF})). 
In these plots only the ``best" fits are shown which correspond to the numbers in bold in Tab. \ref{tab:fit-param-flow-ratio}.  
}
\label{fig:ratios}
\end{figure*}

\begin{table}[htbp]
\centering
\begin{tabular}{c|c|c|c|c|c|c}
\hline
\hline
Discretization
& \(\beta\)
& \(m_s/m_l\)
& \(\sqrt{t_0}/w_0\)
& \(\sqrt{t_2}/w_2\)
& \(\sqrt{t_0}/\sqrt{t_2}\)
& \(w_0/w_2\) \\
\hline
\multirow{4}{*}{Clover}
& \multirow{4}{*}{7.030}
& 5
& 0.8926(10)
& 0.8350(11)
& 1.3087(9)
& 1.2246(11) \\
\cline{3-7}
& & 20
& 0.8808(4)
& 0.8254(7)
& 1.3144(5)
& 1.2316(6) \\ \cline{3-7}
& & $27^*$
& 0.8798(5)
& 0.8245(8)
& 1.3149(5)
& 1.2322(7) \\ \cline{3-7}
& & $27^\dagger$
& 0.8804(5)
& 0.8251(8)
& 1.3146(5)
& 1.2318(6) \\
\cline{2-7}
& \multirow{4}{*}{7.825}
& 5
& 0.8490(40)
& 0.7631(25)
& 1.3602(17)
& 1.2221(39) \\
\cline{3-7}
& & 20
& 0.8481(45)
& 0.7599(29)
& 1.3623(22)
& 1.2219(33) \\ \cline{3-7}
& & $27^*$
& 0.8478(48)
& 0.7597(31)
& 1.3625(24)
& 1.2218(38) \\ \cline{3-7}
& & $27^\dagger$
& 0.8480(46)
& 0.7599(30)
& 1.3624(22)
& 1.2218(35) \\
\hline
\multirow{8}{*}{Improved}
& \multirow{4}{*}{7.030}
& 5
& 0.8484(10)
& 0.7555(12)
& 1.3669(12)
& 1.2176(11) \\
\cline{3-7}
& & 20
& 0.8366(4)
& 0.7461(8)
& 1.3733(5)
& 1.2246(7) \\ \cline{3-7}
& & $27^*$
& 0.8356(5)
& 0.7452(8)
& 1.3738(5)
& 1.2252(7) \\ \cline{3-7}
& & $27^\dagger$
& 0.8363(4)
& 0.7458(8)
& 1.3734(5)
& 1.2248(7) \\
\cline{2-7}
& \multirow{4}{*}{7.825}
& 5
& 0.8370(40)
& 0.7389(24)
& 1.3818(18)
& 1.2194(38) \\
\cline{3-7}
& & 20
& 0.8361(59)
& 0.7348(37)
& 1.3845(23)
& 1.2170(62) \\ \cline{3-7}
& & $27^*$
& 0.8360(66)
& 0.7345(38)
& 1.3848(26)
& 1.2170(66) \\ \cline{3-7}
& & $27^\dagger$
& 0.8361(61)
& 0.7347(37)
& 1.3846(24)
& 1.2170(63) \\
\hline
\hline
\end{tabular}
\caption{Comparison of ratios of gradient-flow scales at the two common
couplings \(\beta=7.030\) and \(\beta=7.825\) for \(m_s/m_l=5\) and
\(m_s/m_l=20\), using both the clover and improved definitions of the
energy density. The results for $m_s/m_l = 27$ are obtained via extrapolation, using fits to the data at $m_s/m_l = 5$ and $20$. The extrapolations are based on an ansatz that is linear in $m_l$ (denoted by a superscript $*$) and quadratic in $m_l$ (denoted by a superscript $\dagger$), respectively.}
\label{tab:flow-ratios-ms5-ms20}
\end{table}

We determine the gradient flow scales $t_i$ and $w_i$, $i=0,2$ through Eq.~(\ref{eq:fixpoint}) using clover or improved discretization for the action density. The numerical results for the action density, $E$ and the derivative of $\tauf E$ as function of the flow time are shown in Fig. \ref{fig:t2E} for improved
discretizations. The results for clover discretization look similar.  We interpolate the numerical results on the action density in flow time using splines. From the splines it is straightforward to
estimate the derivative of $\tauf E$ as well as to evaluate the gradient flow scales. 
Our results for the gradient flow scales in lattice units are  summarized in Tab. \ref{tab:flowscales20} and \ref{tab:flowscales5} in Appendix \ref{app:flowscales}. 
The $t_2$ and $w_2$ scales have
smaller statistical errors compared to $t_0$ and $w_0$ scales, which is advantageous for large values of $\beta$. For $\beta\ge 6.74$ the gradient flow scales are determined for
both clover and improved discretization. There are differences between the gradient flow
scales obtained using clover and improved discretizations schemes. These differences are
the largest for $t_2$ and the smallest for $w_0$. Furthermore, as expected, the differences between the gradient flow scales obtained using clover and improved discretizations decrease
with increasing $\beta$.

As mentioned in the previous section, for $\beta \ge 8.0$ the individual Monte-Carlo streams 
correspond to frozen topology. Fortunately, we have streams that correspond to several values
of topological charge. One may hope that the gradient flow scales being short distance quantities, i.e.
being determined at relatively small values of the flow time, will not be affected
much by the frozen topology. It has been shown that this is the case for the potential scales
$r_1$ and $r_2$ determined at these values of $\beta$ \cite{Bazavov:2017dsy}.
We perform a similar analysis for the gradient flow scale in Appendix \ref{app:topo-analysis},
where we show that the values of the gradient flow scale obtained using Monte-Carlo streams 
with different values of the topological charge agree within the statical errors.
Thus at present level of statistical accuracy we do not see any effects of topological freezing
on the values of the gradient flow scales.

We performed Allton fits of the gradient flow scales $\sqrt{t_0}/a$ and $w_0/a$ obtained from the improved discretization for
$m_s/m_l=20$ and $6.740\leq \beta \leq 7.825$.
The parameters of these fits 
are summarized in Tab. \ref{tab:fit-scale-setting}
and shown in Fig.~\ref{fig:flow-scales-setting}.
While the $\chi^2$/d.o.f. of the fit is large
due to the tiny errors on the gradient flow scale, overall the fits capture the $\beta$ dependence of the scale
quite well. We also performed Allton fits for $\sqrt{t_0}/a$ and $w_0/a$ obtained for $m_s/m_l=5$ with improved discretization.
The results of these fits are also shown in Fig.~\ref{fig:flow-scales-setting} and summarized in Tab. \ref{tab:fit-scale-setting}. Note that the Allton-type parametrizations shown here are not used in the subsequent analysis to determine gradient-flow scales at values of \(\beta\) where no simulation data are available. The purpose of these parametrizations is to provide a convenient estimate of the lattice spacing at intermediate values of \(\beta\) within the simulated range, and possibly also to guide estimates at somewhat larger values of \(\beta\) beyond the present range.

Next we determine the ratio of different gradient flow scales: $\sqrt{t_0}/w_0$, 
$\sqrt{t_2}/w_2$, $\sqrt{t_0}/\sqrt{t_2}$ and $w_0/w_2$ using the \(m_s/m_l=20\) ensembles in the range \(6.740\leq \beta\leq 7.825\). Since these scales are measured on the same ensembles, nontrivial correlations naturally arise among them. To properly account for these correlations, we evaluate each scale separately using the same bootstrap samples. In addition, autocorrelations are present within each measured scale. These are handled by binning the data with a bin size equal to twice the integrated autocorrelation time, $\tau_{\mathrm{int}}$, before performing the bootstrap analysis. Here, $\tau_{\mathrm{int}}$ is taken to be the larger of the values obtained from the numerator and the denominator entering the ratio.
These ratios are shown in Fig. \ref{fig:ratios} as function of the lattice spacing in units of $w_0$. 
As one can see from the figures
the results from clover discretization show more
pronounced dependence on the lattice spacing 
compared to the ones from improved discretization. Interestingly, the ratio $w_0/w_2$ shows very little lattice spacing dependence for improved discretization as can be seen from bottom right panel of Fig. \ref{fig:ratios}.

For clover discretization we expect that these ratios scale as $a^2$. For improved discretization we expect the lattice spacing dependence to be proportional to $\alpha_s a^2$. We see that indeed the $a^2$ dependence of these ratios is much reduced
compared to clover discretization. We performed continuum extrapolations of these ratios
which are shown as bands in Fig. \ref{fig:ratios}. For clover discretization we performed
$a^2$ extrapolations, while for improved discretization we performed $\alpha_b a^2$ extrapolations, with $\alpha_b$ being the boosted lattice gauge coupling
$\alpha_b=g_0^2/(4 \pi u_0^4)=10/(4 \pi \beta u_0^4)$. Here $u_0^4$ stands for the average plaquette taken from \cite{Bazavov:2019qoo}. For the improved discretization we also consider the $a^2$ extrapolations as an effective description of the $\alpha_b a^2$ extrapolations. 
The continuum-extrapolated results 
for these ratios show some sensitivity to the choice of the range of lattice spacing used in the fits.
Furthermore, the lattice spacing dependence of
these ratios may not be described by the above simple
forms when the lattice spacing is too large. In addition, the range of lattice spacing that can be used for the continuum extrapolations could be different
to clover and improved discretization.
To examine the systematic uncertainties stemming from these, we perform the continuum extrapolations by excluding different number of coarser lattices. The fit results are summarized in Tab. \ref{tab:fit-param-flow-ratio}. We can see that some of the fits can be identified as good while some can not. 
For the good fits the continuum extrapolated values 
do not depend on the $\beta$ range within errors.
To pick out the good fits we adopt the following criteria i) $\chi^2/$d.o.f. should be the closest to 1.0 (keeping one decimal place); ii) if there are two such fits, we take the one that has more number of data points. 
The selected best fits are indicated in boldface in Tab. \ref{tab:fit-param-flow-ratio}. Performing a weighted average of the selected best-fit values, and estimating the systematic uncertainty as the weighted root-mean-square deviation from this average (using the statistical errors as weights), we obtain
\begin{equation}
\begin{split}
\sqrt{t_0}/w_0 &= 0.8291(8)(6),\\
\sqrt{t_2}/w_2 &= 0.7323(9)(37),\\
\sqrt{t_0}/\sqrt{t_2} &= 1.3867(10)(36),\\
w_0/w_2 &= 1.2240(8)(8),
\end{split}
\label{eq:ratio-flow-combined}
\end{equation}
where the first parentheses denote statistical uncertainties and the second denote systematic uncertainties. The averaged results are shown as red data points in Fig. \ref{fig:ratios}, with the error bar being the sum of the statistical uncertainties and the systematic uncertainties in quadrature. The obtained $\sqrt{t_0}/w_0$ is consistent with the $N_f$=2+1+1 estimate from EMT Collaboration \cite{ExtendedTwistedMass:2021qui} 0.82930(65) and the $N_f$=2+1+1 estimate from HPQCD Collaboration \cite{Dowdall:2013rya} 0.835(8). We note that the ratios reported here are obtained from the \(m_s/m_l=20\) ensembles, rather than from ensembles at the physical light-quark mass. The quark-mass dependence of these ratios is expected to be mild, since gradient-flow scales are gluonic observables, and the use of ratios should further suppress the common mass dependence. To estimate the possible size of this effect, we compare the results from the \(m_s/m_l=20\) and \(m_s/m_l=5\) ensembles at the two lattice spacings where both data sets are available, \(\beta=7.030\) and \(\beta=7.825\), as summarized in Tab.~\ref{tab:flow-ratios-ms5-ms20}. Using the $m_s/m_l=20$ results as the baseline, we find that the absolute relative differences are somewhat larger at $\beta=7.030$. At the finer lattice spacing, $\beta=7.825$, they are approximately $0.5\%$ for $\sqrt{t_2}/w_2$ and of order $0.1\%$ for the other ratios, comparable to the combined uncertainties quoted in Eq.~(\ref{eq:ratio-flow-combined}). We further fit the data at $m_s/m_l=5$ and $20$ using ans\"atze that are linear and quadratic in $m_l$, denoted by the superscripts $*$ and $\dagger$, respectively, in Tab.~\ref{tab:flow-ratios-ms5-ms20}, and extrapolate to the physical point, $m_s/m_l=27$. The difference between the two extrapolations is negligible compared with their statistical uncertainties. Moreover, the extrapolated values are very close to those obtained at $m_s/m_l=20$, with the differences covered by the statistical errors.

\subsection{The relation of the gradient flow scales to the potential scales}
\begin{figure*}[tbh]
\centerline{
\includegraphics[width=0.5\textwidth]{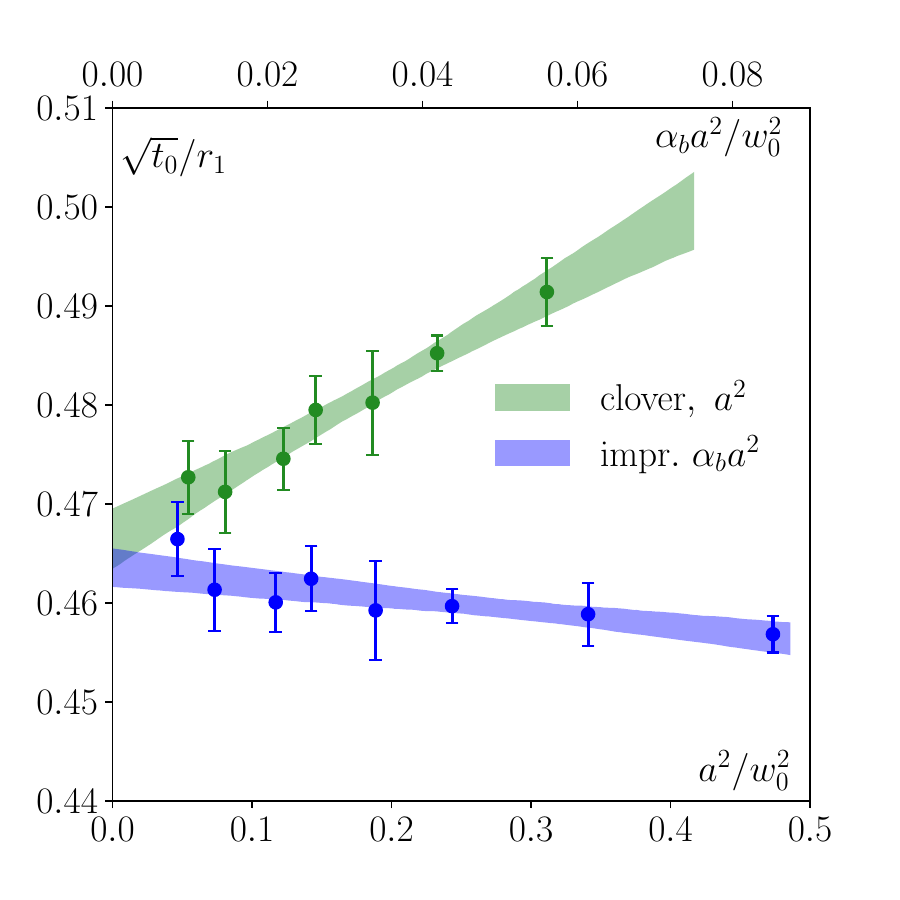}
\includegraphics[width=0.5\textwidth]{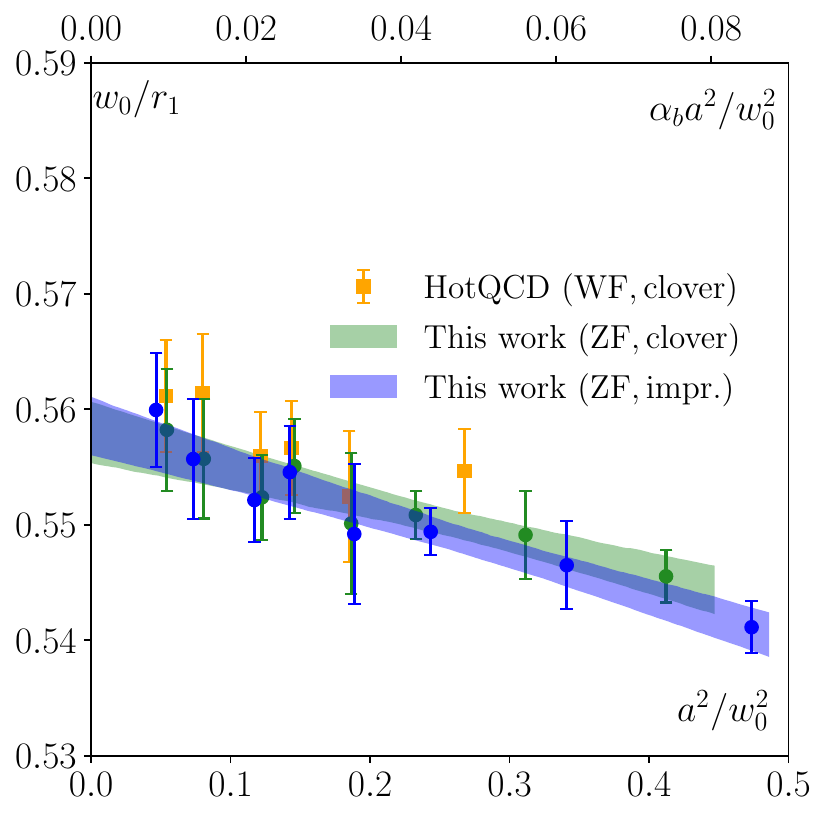}
}
\caption{Continuum extrapolation of the ratio of the flow scales to the $r_1$ scale. Left: $\sqrt{t_0}/r_1$. Similar to before, the discretization effects are significant in this case. After applying the $\mathcal{O}(a^2)$ improvement to the field strength tensor, the slope of the extrapolation (using Ansatz linear in $a^2$) changes sign. However, in the continuum limit, the two discretizations give consistent results. Right: $w_0/r_1$. ``WF" stands for Wilson flow and ``ZF" for Zeuthen flow. We include results from the HotQCD Collaboration (red points) \cite{HotQCD:2014kol} using Wilson flow and clover-discretized action density for comparison. It can be observed that for $w_0$, the discretization effects are so mild that the two discretizations give almost the same results. Our results using Zeuthen flow are slightly smaller than the HotQCD Collaboration's previous results using Wilson flow. For both panels, we present only the $a^2$ fits for the clover discretization
and the $\alpha_b a^2$ fits for the improved discretization. The quantity \(a^2/w_0^2\) shown on the abscissa is computed using the corresponding discretization. For the improved discretization, one should refer to the upper abscissa, where the factor \(\alpha_b\) has been included.}
\label{fig:flow-r1}
\end{figure*}

\begin{figure*}[tbh]
\centerline{
\includegraphics[width=0.5\textwidth]{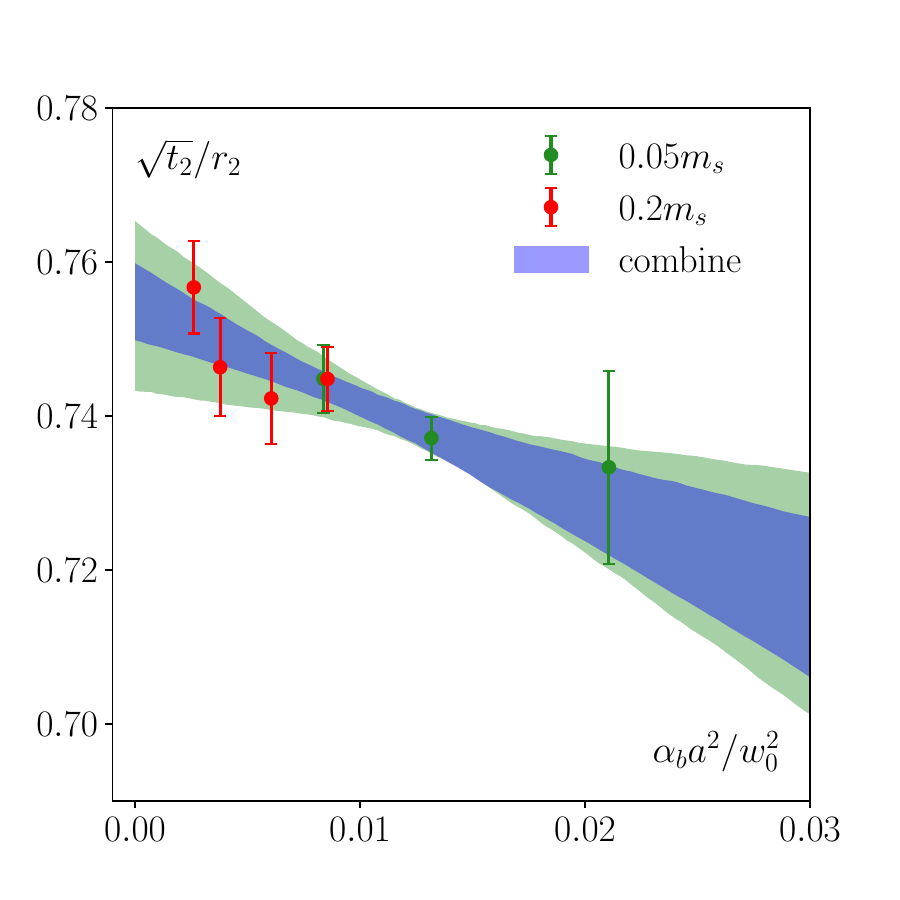}
\includegraphics[width=0.5\textwidth]{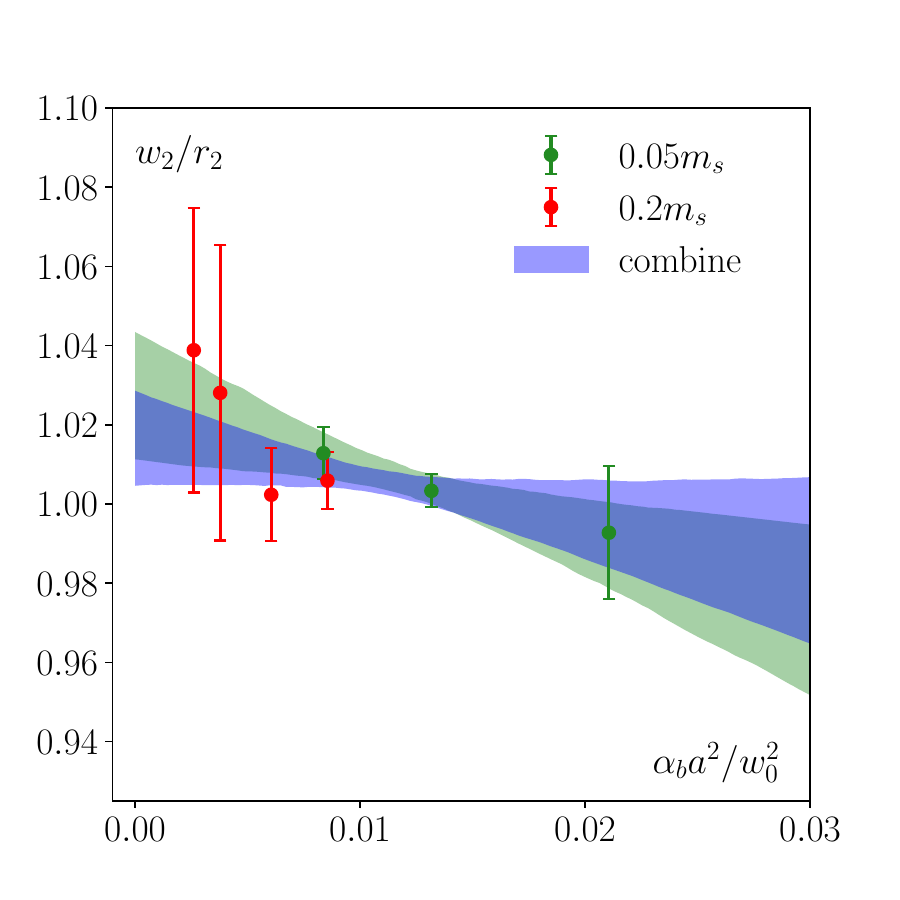}
}
    \caption{The ratio of the gradient flow slales $\sqrt{t_2}$ and $w_2$ to $r_2$ scale as
    function of the lattice spacing. 
    }
    \label{fig:flow-r2}
\end{figure*}
As mentioned in the introduction, theory scales based on the static quark-antiquark potential $V(r)$ are often used in lattice QCD calculations. These scales are defined as
\begin{equation}
    \left . r^2 \frac{d V(r)}{dr}\right |_{r=r_i}=b_i,~i=0,1,2. 
\end{equation}
The choices $b_0=1.65$, $b_1=1$ and $b_2=0.5$ define the $r_0$ (Sommer) \cite{Sommer:1993ce}, $r_1$ scale \cite{Bernard:2000gd} and $r_2$ scale \cite{Bazavov:2017dsy}, respectively. In this subsection we express 
the gradient flow scales in terms of $r_1$ and $r_2$ scales. Since the discretization effects in the static
potential are expected to be small at distances around $r_1$ and $r_2$, the ratio of the gradient flow scales
and the potential scales can give some insight into the lattice artifacts in the determination of the gradient flow scales.
Based on the $6.740\leq \beta \leq 7.825$ lattices listed in the first block of Tab. \ref{tab:param-zeroT}, our results for the ratios $r_1/\sqrt{t_0}$ and $r_1/w_0$ are shown in Fig. \ref{fig:flow-r1}, where we use the values of $r_1/a$ determined in Ref.~\cite{HotQCD:2014kol}.
For the latter we compare our results with the results of HotQCD Collaboration \cite{HotQCD:2014kol} obtained using Wilson flow and clover-discretized action density for comparison. Our results agree with the HotQCD results within errors. As
one can see from the figure of $r_1/\sqrt{t_0}$, 
improved discretization scheme leads to much smaller cutoff effects. On the other hand
for $r_1/w_0$ there is no visible difference between the two discretization schemes and 
also Wilson flow leads to similar results. This suggests that lattice artifacts
are smaller in the $w_0$ scale compared to the $t_0$ scale. We performed continuum extrapolations for $r_1/\sqrt{t_0}$
and $r_1/w_0$ for both clover and improved discretization. For clover discretization we assume $a^2$-dependence and exclude the $\beta=6.740$ ensemble to gain better fits. For improved action density we further
perform $\alpha_b a^2$ extrapolations. All the continuum extrapolations of these ratios agree within the estimated
errors, see c.f. Fig. \ref{fig:flow-r1}. Using improved action density and $\alpha_b a^2$ extrapolation we obtain:
\begin{equation}
r_1/\sqrt{t_0}=2.1572(91),~ r_1/w_0 =1.7904(81).
\label{eq:r1flow}
\end{equation}
Our result for $r_1/w_0$ could be compared with HotQCD result,
$r_1/w_0=1.7797(67)$ for $N_f=2+1$ \cite{HotQCD:2014kol} and the HPQCD result, $r_1/w_0=1.789(26)$ for $N_f=2+1+1$ \cite{Dowdall:2013rya}. 
We see that all results for $r_1/w_0$ agree within errors.
Recently the $r_1/a$ scale has been determined in 2+1+1 flavor QCD using HISQ action by TUMQCD collaboration for physical light quark masses and
lattice spacings $a=0.04,~0.06,~0.09,~0.12$ and $0.15$ fm \cite{Brambilla:2022het}. The gradient flow scales   $\sqrt{t_0}/a$ and $w_0/a$ 
have also been determined on these lattices \cite{MILC:2015tqx,Brown:2018jtv,Bazavov:2025mao}. In the most recent study also improved discretization for
the gauge action density was used to determine $w_0$. Using the results on $t_0/a$ from Ref. \cite{Brown:2018jtv} and the most recent results 
on $w_0/a$ from Ref. \cite{Bazavov:2025mao}
we estimated the continuum values of $r_1/\sqrt{t_0}$ and $r_1/w_0$ assuming that discretization effects scale as $\alpha_b a^2$:
\begin{equation}
r_1/\sqrt{t_0}=2.151(10),~ r_1/w_0=1.7713(74),~N_f=2+1+1.
\label{eq:r1flow_4f}
\end{equation}
Here we use the values of $w_0$ and $t_0$ computed in Ref.~\cite{MILC:2015tqx}, where a perturbative partial tree-level improvement of the observables was adopted~\cite{Fodor:2014cpa}. Neglecting the residual tree-level discretization effects, namely terms of $O(a^n)$ with $n=4$--$8$, the remaining leading cutoff effects are expected to scale as $O(\alpha_s a^2)$, same as in the improvement employed in this work. The resulting value of $r_1/\sqrt{t_0}$ is consistent with our 2+1 flavor determination, whereas the value of $r_1/w_0$ is slightly smaller than the corresponding 2+1 flavor result. 
We note that while our analysis was performed at $m_l/m_s=1/20$ instead of the physical value of $1/27$ the small deviation from the physical light quark mass
has no visible effect at present statistical accuracy. As it was shown above for the gradient flow scales the difference between $m_l/m_s=1/20$ and $m_s/m_l=1/27$ is smaller than the statistical
errors. In Ref. \cite{Bazavov:2017dsy} the $r_1$ scale was calculated at $\beta=7.030$ and $\beta=7.825$ for $m_l/m_s=20$ and $m_l/m_s=5$ and the difference was about twice the estimated error 
and smaller than the estimated error. Therefore, we expect that for $r_1$ scale the difference between $m_l/m_s=1/20$ and $m_l/m_s=1/27$ should be negligible at present.

\begin{figure*}[tbh]
\centerline{
\includegraphics[width=0.33\textwidth]{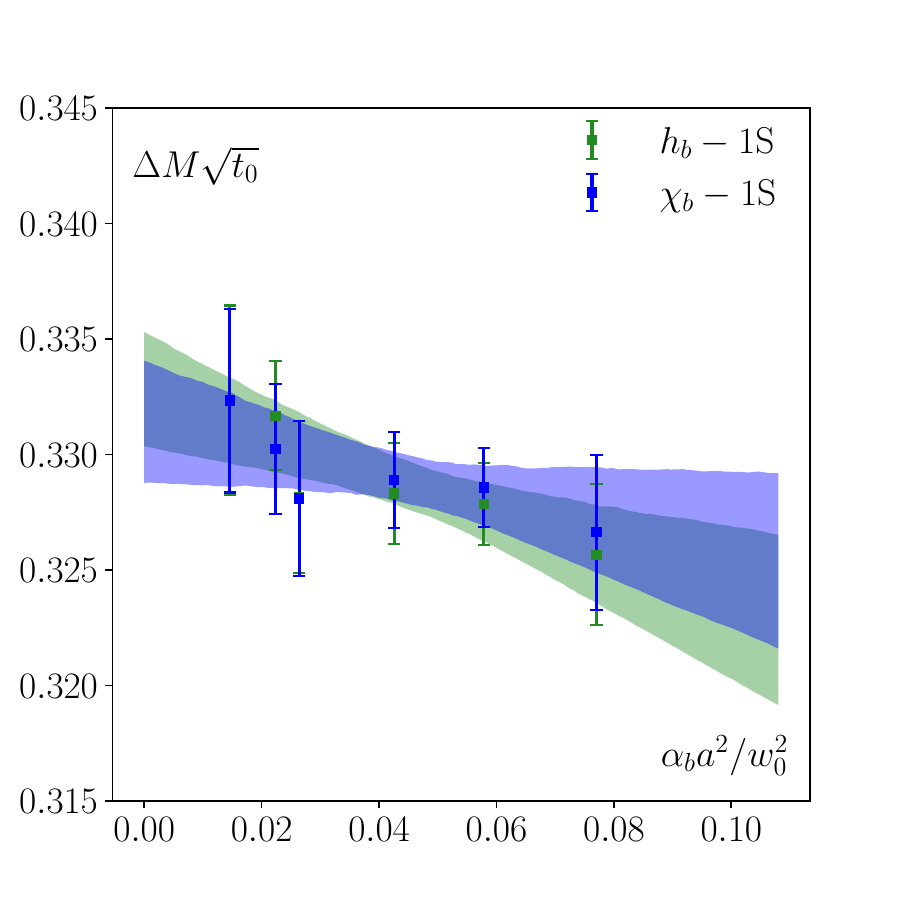}
\includegraphics[width=0.33\textwidth]{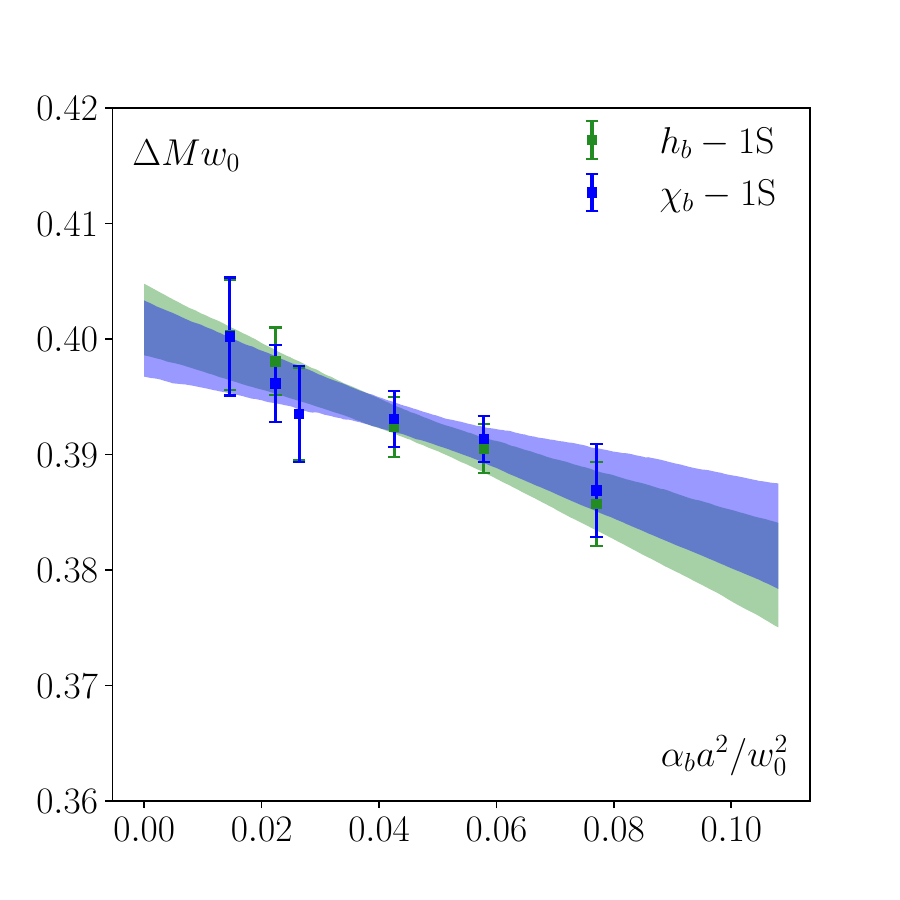}
\includegraphics[width=0.33\textwidth]{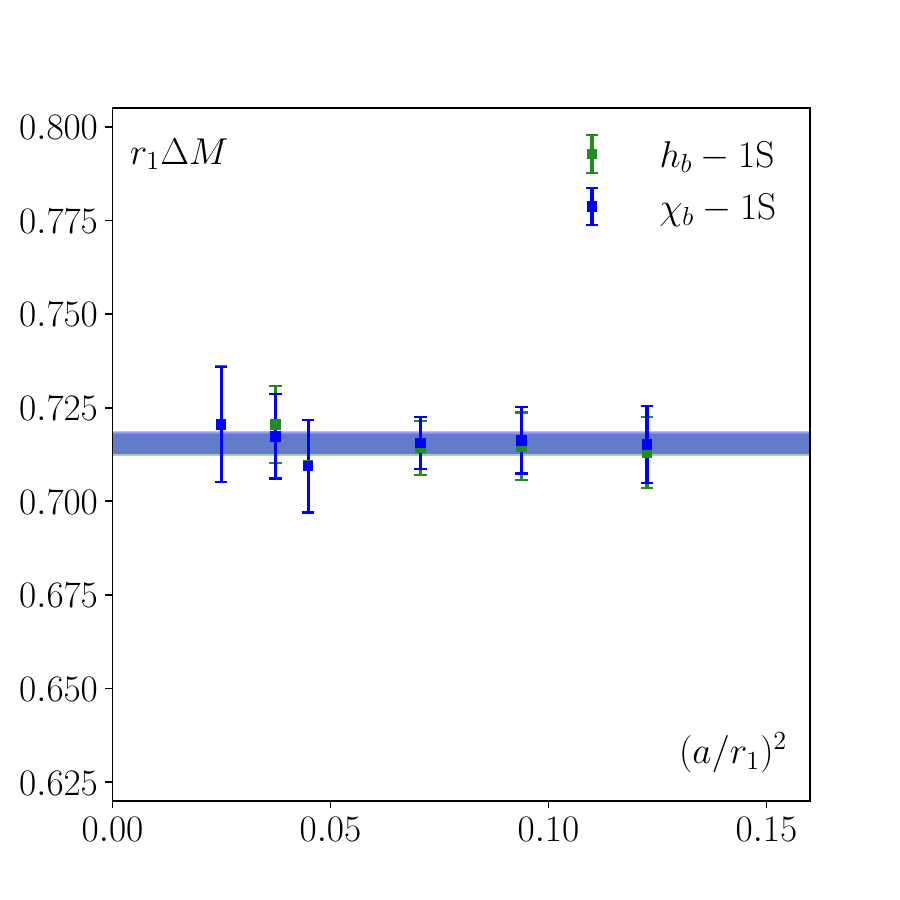}
}
\centerline{
\includegraphics[width=0.33\textwidth]{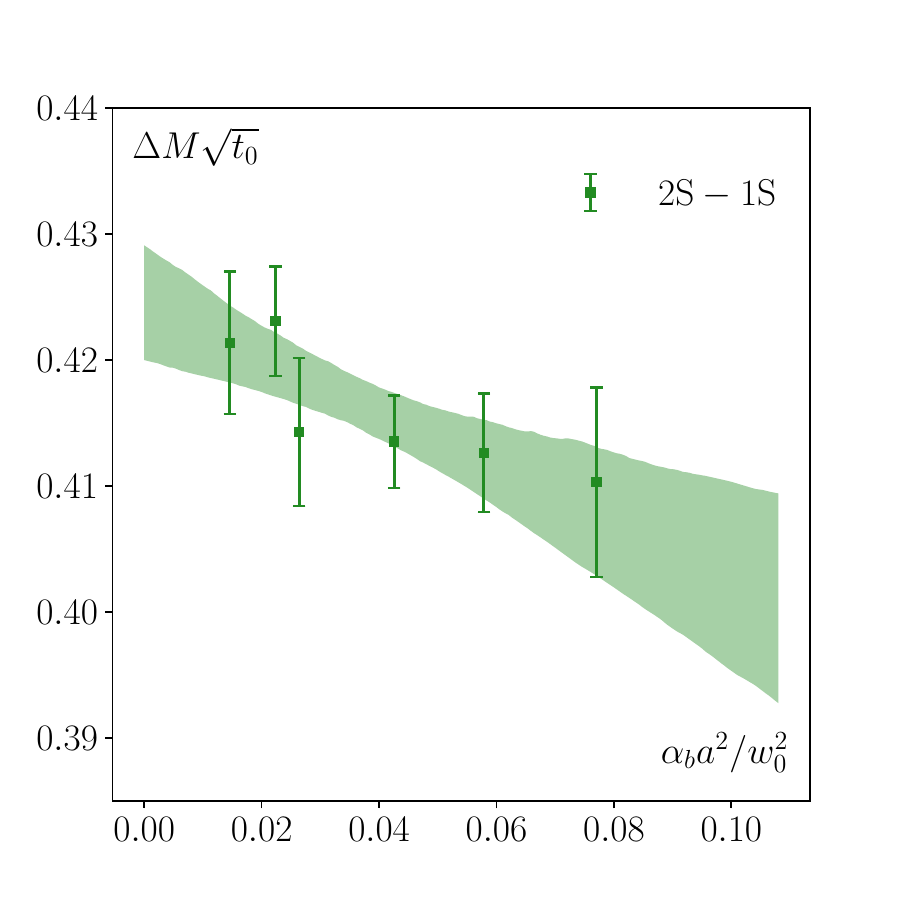}
\includegraphics[width=0.33\textwidth]{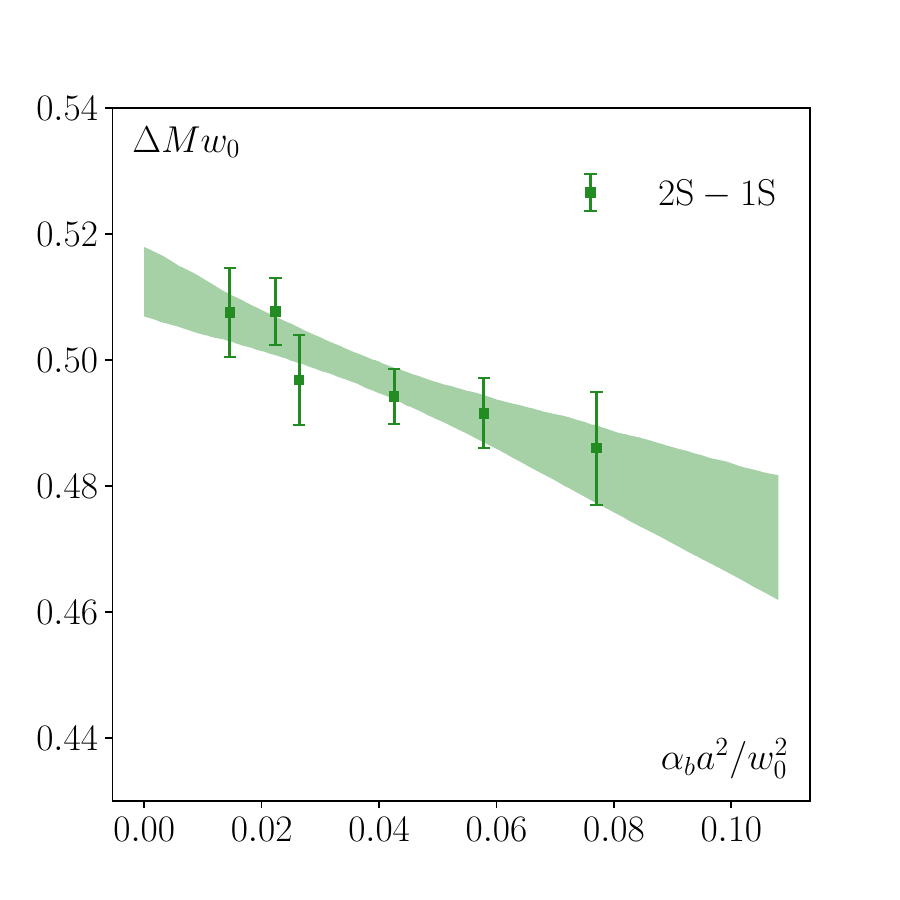}
\includegraphics[width=0.33\textwidth]{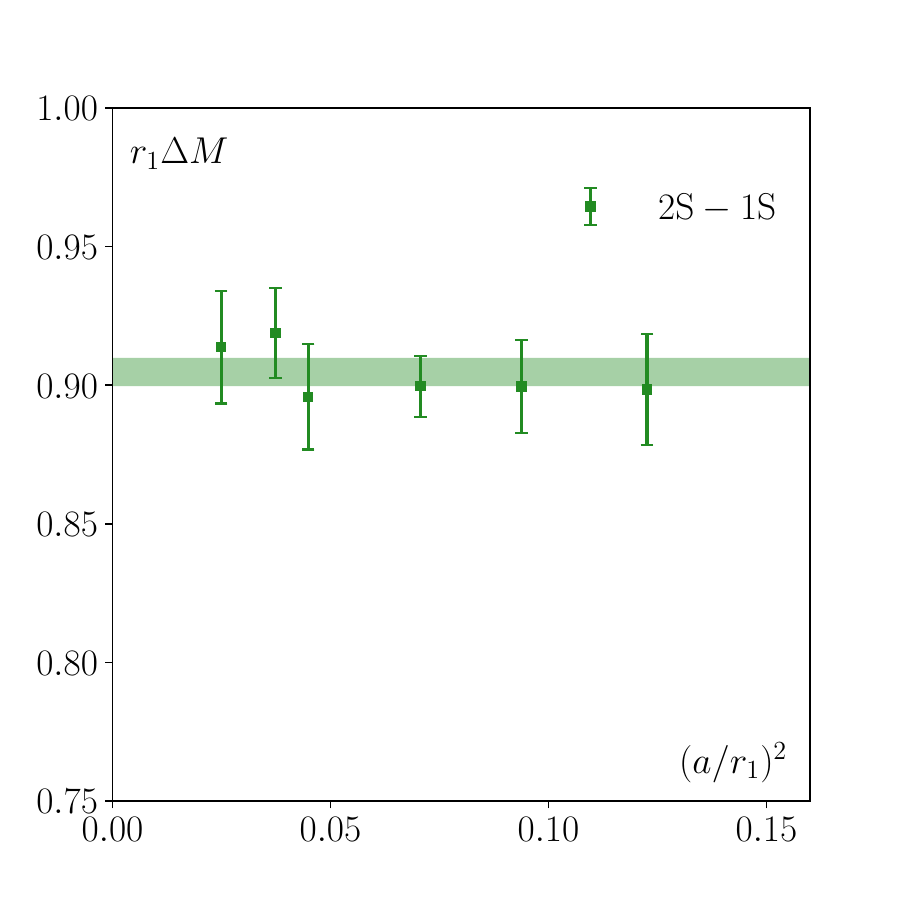}
}
\caption{The continuum extrapolation of the products $\Delta M \sqrt{t_0}$ (left), $\Delta M w_0$ (middle) and $\Delta M r_1$ (right) via $\Delta M_{\mathrm{1P-1S}}$ (top) and $\Delta M_{\mathrm{2S-1S}}$ (bottom). The mass differences $\Delta M$ are taken from Refs. \cite{Ding:2025fvo,Larsen:2019zqv}. The bands represent the continuum extrapolation. }
\label{fig:theory-from-dm}
\end{figure*}

In Fig. \ref{fig:flow-r2} we show the ratio of the gradient flow scales $\sqrt{t_2}$
and $w_2$ to $r_2$. 
We use data for $r_2/a$ obtained at $\beta=7.373$, 7.596 and 7.825, $m_s/m_l=20$ and
$\beta=7.825$,~8.000, 8.200 and 8.400 for $m_s/m_l=5$ computed in  Ref.~\cite{Bazavov:2017dsy}. The light quark mass effects in this ratio appear to be quite small. We performed continuum extrapolations for these ratios using $\alpha_b a^2$ form.
These continuum extrapolations are shown in Fig. \ref{fig:flow-r2} both including and excluding the $m_s/m_l=5$ lattices. Using all the available data we
obtain:
\begin{equation}
\sqrt{t_2}/r_2=0.7548(50),~ w_2/r_2=1.0166(120).
\end{equation}
Excluding the $m_s/m_l=5$ lattices does not change the above
result but only increases the errors slightly.

\section{Gradient flow scales in physical units}\label{results}

To determine the gradient flow scales in physical units, one computes the
dimensionless products of these scales with suitable physical quantities such
as hadron masses, quarkonium mass splittings, or meson decay constants at several
lattice spacings, and extrapolates these products to the continuum limit.  
Using the experimentally measured values of the corresponding physical
quantities, the gradient flow scales can then be expressed in physical units.
In the following, we discuss their determination using bottomonium
mass splittings, pseudo-scalar meson decay constants, and the $\phi$-meson mass.

\begin{table*}[tbh]
\centering
\setlength{\tabcolsep}{3pt}
\scriptsize
\begin{tabular}{ccccccc}
\hline \hline
  & $\Delta M \sqrt{t_0}$ & $\Delta M w_0$ &  $\Delta M r_1$ & $\sqrt{t_0}$ [fm] & $w_0$ [fm] & $r_1$ [fm] \\ \hline
$h_b-\mathrm{1S}$ &  0.33282(247) & 0.40168(310) & 0.71511 (297) & 0.14455(108)(25)(16) & 0.17445(132)(31)(19) & 0.3106(13)(5)(3) \\
$\chi_b-\mathrm{1S}$ &  0.33141(264) & 0.40004(330) & 0.71568(299) & 0.14376(115)(13)(16) & 0.17350(140)(15)(19) & 0.3104(13)(3)(3) \\
$\mathrm{2S}-\mathrm{1S}$ &  0.42455(456) & 0.51242(551) & 0.90479(511) & 0.14635(152)(28)(13) & 0.17664(184)(34)(15) & 0.3120(17)(6)(3) \\
\hline \hline
\end{tabular}
\caption{The bottomonium mass splittings obtained from PDG \cite{ParticleDataGroup:2018ovx} and the continuum-extrapolated values of $\Delta M\sqrt{t_0}$, $\Delta M w_0$ and  $\Delta M r_1$. The physical values of the extracted theory scales $\sqrt{t_0}$,  $w_0$ and $r_1$ are also provided. The uncertainties in the products of the bottomonium mass splittings and the theory scales are purely statistical. For the quoted theory scales, the first errors denote the statistical uncertainties, the second errors are propagated from the uncertainties in the \(h_b\), \(\chi_b\), and \(2S\) masses, respectively, and the third errors are propagated from the uncertainty in the \(1S\) mass.}
\label{tab:theory-from-dm}
\end{table*}

\subsection{Determination of the theory scales from bottomonium mass splitting}
\label{sec:bottomonium-mass-splitting}
Bottomonium mass splittings have been used in the determination of the theory scales in the past \cite{Gray:2005ur,Davies:2009tsa,HPQCD:2011qwj}.
In this sub-section we discuss the determination of the theory scales using bottomonium splitting.
We use the lattice results on energy levels of different bottomonium states calculated with nonrelativistic QCD (NRQCD) 
on 2+1 flavor HISQ gauge configurations in Ref. \cite{Larsen:2019zqv} for $\beta=6.740, 6.880, 7.03, 7.28, 7.596$ and in Ref. \cite{Ding:2025fvo} for $\beta=7.373$.
In NRQCD absolute value of bottomonium masses are difficult to determine and one usually considers bottomonium mass splittings instead.
Since the bottomonium mass differences calculated in NRQCD can be affected by missing higher order relativistic corrections as well as missing radiative corrections to the parameters of NRQCD Lagrangian, it
is important to consider mass differences that minimize these effects. 
The missing corrections in NRQCD Lagrangian mostly affect the mass splittings which are sensitive to spin-spin and spin-orbit interactions. To get rid of the effects of spin-spin interactions we consider the spin averaged 
1S and 2S bottomonium masses, $M_{\mathrm{1S}}=(M_{\eta_b(\mathrm{1S})}+3M_{\Upsilon(\mathrm{1S})})/4$
and $M_{\mathrm{2S}}=(M_{\eta_b(\mathrm{2S})}+3M_{\Upsilon(\mathrm{2S})})/4$
and their difference. 
To minimize the effect to spin-orbit interactions we consider the difference between the mass of $h_b(1P)$ state and 
$M_{\mathrm{1S}}$ as well as the difference between the spin-averaged 
$\chi_b(1P)$ mass defined
as $M^{\chi_b}_{\mathrm{1P}}=(1M_{\chi_{b0}(\mathrm{1P})}+3M_{\chi_{b1}(\mathrm{1P})}+5M_{\chi_{b2}(\mathrm{1P})})/9$ and
$M_{\mathrm{1S}}$.
We denote the above mass differences as $\Delta M$ with additional labels $2S-1S$, $h_b-1S$ and $\chi_b-1S$.
The above mass difference have very small dependence on the light quark mass if $m_{\pi}<300$ MeV \cite{Dowdall:2013rya,Meinel:2010pv}. Therefore,
the fact that the light quark mass is $m_s/20$ in our calculations instead of the physical value of $m_s/27$ should have no effect on the scale determination.

Similar to previous section, we calculate the dimensionless products $\Delta M\sqrt{t_0}$, $\Delta M w_0$ and $r_1\Delta M$ present on the same ensembles, and then extrapolate them to the continuum limit. The continuum extrapolation of these dimensionless products are shown in Fig.~\ref{fig:theory-from-dm}.
For the continuum extrapolation of $\Delta M\sqrt{t_0}$ and $\Delta M w_0$ we use an Ansatz linear in $\alpha_b a^2/w_0^2$.
Interestingly enough, the combination  $r_1\Delta M$ shows no dependence on the lattice spacings within errors, see Fig.~\ref{fig:theory-from-dm}.
Therefore, we fit this combination by a constant. The continuum fits are shown as bands in Fig.~\ref{fig:theory-from-dm}.

\begin{figure*}[tbh]
\centerline{
\includegraphics[width=0.5\textwidth]{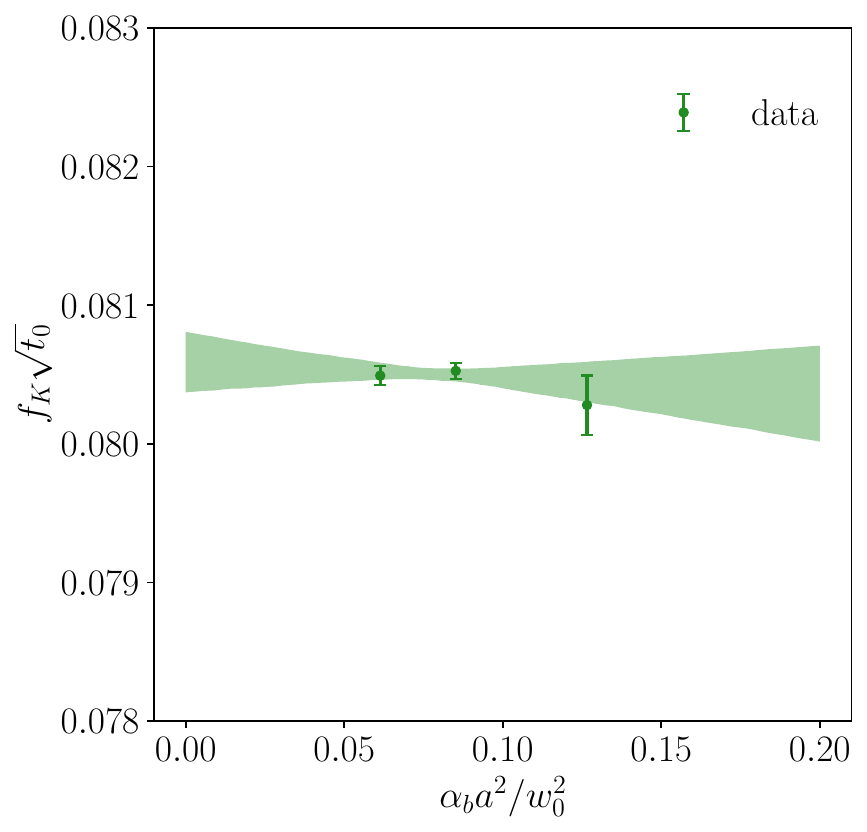}
\includegraphics[width=0.5\textwidth]{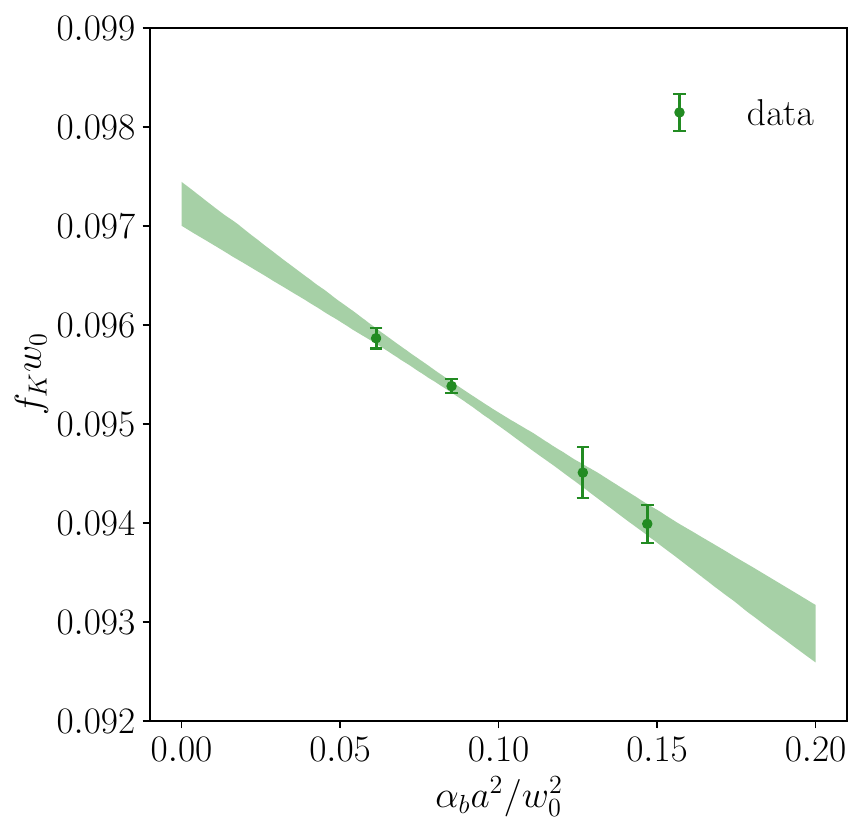}
}
\centerline{
\includegraphics[width=0.5\textwidth]{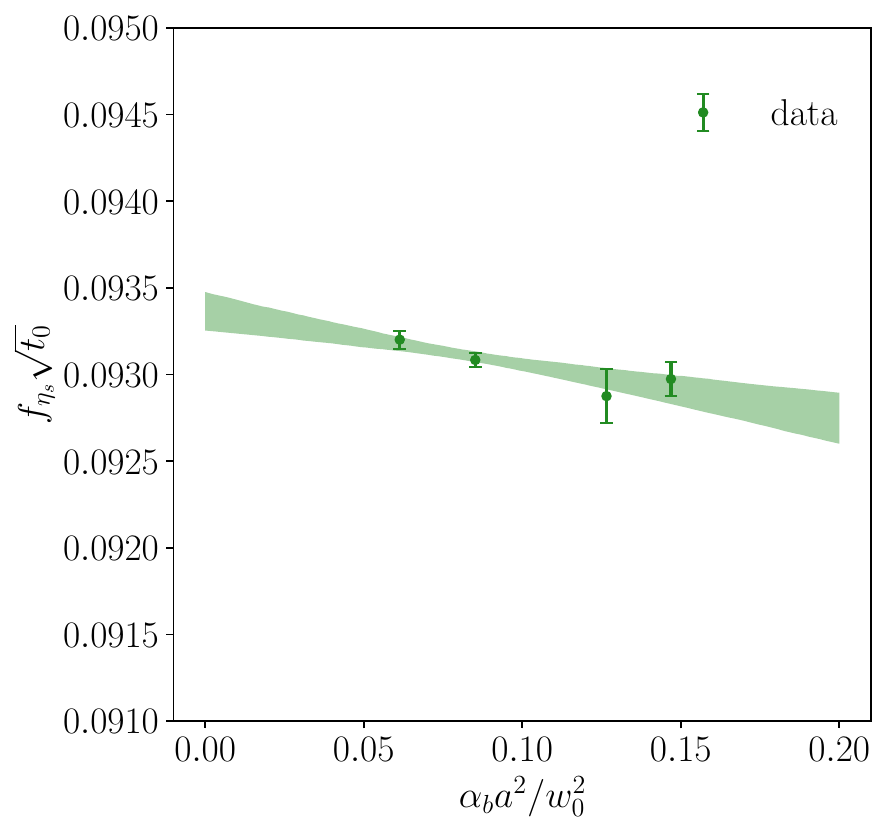}
\includegraphics[width=0.5\textwidth]{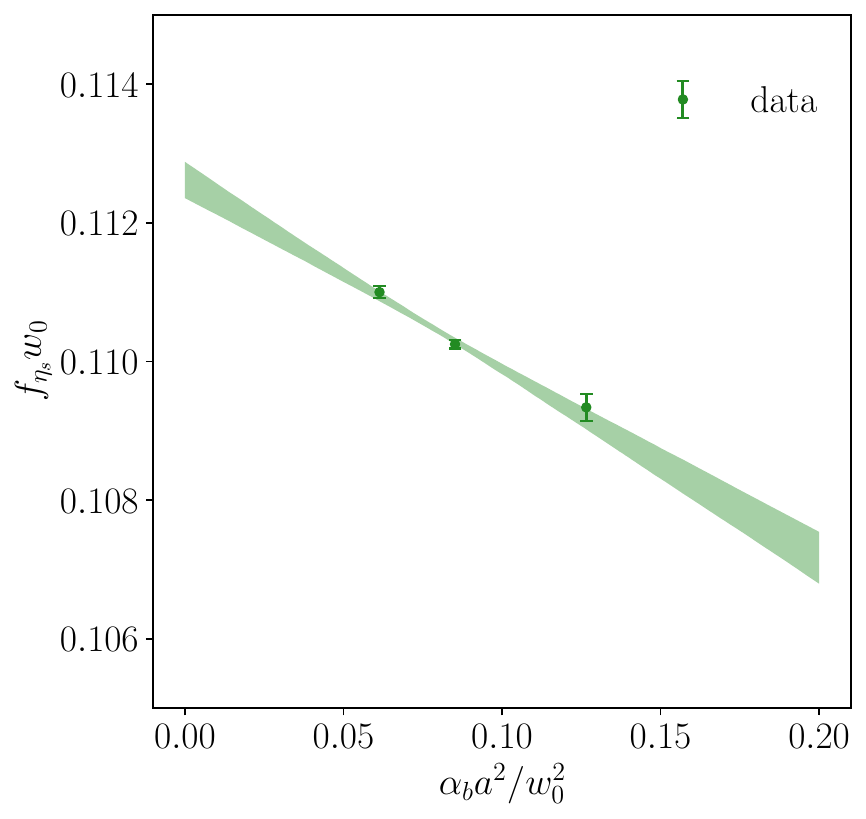}
}
\caption{
Continuum extrapolation of $\sqrt{t_0} f_K$ (top left), $w_0 f_K$ (top right), $\sqrt{t_0} f_{\eta_s}$ (bottom left) and $w_0 f_{\eta_s}$ (bottom right) using the $\beta$=6.575, 6.740
 and 6.880 lattices.}
\label{fig:decay-flow}
\end{figure*}

\begin{table*}[tbh]
\centering
\begin{tabular}{c|c|c|c}
\hline \hline 
& via $f_{\eta_s}$ & via $f_K$ & via $m_\phi$ \\
\hline
$\sqrt{t_0}$ [fm] &  0.14354 (17) (95) (21) & 0.14443 (39) (65) (35) & 0.14461 (43) (6) (67) \\ 
$w_0$ [fm] & 0.17315 (40) (114) (46) & 0.17425 (40) (78) (15) & 0.17389 (51) (7) (59) \\
$r_1$ [fm] & 0.3105 (37) (21) (7) & 0.3131 (37) (14) (1)  & 0.3122 (31) (1) (10)\\
\hline \hline 
\end{tabular}
\caption{Theory scales determined via $f_{\eta_s}$, $f_K$ and $\phi$ meson mass. The first error stands for the statistical error. The second error represents the error propagated from the cited decay constants and meson masses. The third error is the systematic error of the continuum extrapolation, which is estimated using the difference between the extrapolated results obtained by including and excluding the coarsest lattice. 
}
\label{tab:flow-from-decay-womass}
\end{table*}

Using the values of $\Delta M$ taken from PDG \cite{ParticleDataGroup:2018ovx}, we can convert the products to extract the theory scales in physical units. The obtained values of the products and the theory scales are listed in Tab. \ref{tab:theory-from-dm}. 
We note that the mass splittings between different spin-averaged bottomomonium states have systematic errors
due to the use of NRQCD. The estimates of these errors vary depending on the details of the lattice setup and the states considered and 
lie between 0.2\% $\sim$ 1.2\% \cite{Dowdall:2013rya,Meinel:2010pv}. We assign a  1\% systematic error to the $h_b-1S$, $\chi_b-1S$ and 
$2S-1S$ splitting and propagate these systematic errors to the corresponding systematic errors on the values of $\sqrt{t_0}$, $w_0$ and $r_1$ in physical units.
Besides, the mass splittings obtained from PDG also carry uncertainties, which propagate into the extracted theory scales and are quoted as the second errors and third errors for $h_b$/$\chi_b$/$2S$ and $1S$, respectively in Tab.~\ref{tab:theory-from-dm}. To determine the central values of the theory scales, we perform a weighted average over the individual determinations, with weights constructed solely from the statistical uncertainties. For the final uncertainty, we first add in quadrature, for each individual determination, its statistical error, the uncertainties propagated from the bottomonium mass splitting, and the (1\%) systematic uncertainty associated with the use of NRQCD. We then take the smallest of these total uncertainties as our final quoted error because these determinations are correlated. We obtain:
\begin{equation}
\begin{split}
\sqrt{t_0} &= 0.14465\ (183)\ \mathrm{fm},\\
w_0 &= 0.17457\ (221)\ \mathrm{fm},\\
r_1 &= 0.3108\ (34)\ \mathrm{fm}.
\end{split}
\label{eq:theory-DM}
\end{equation}
We note that, in the above analysis, there is in principle also a systematic uncertainty associated with the continuum extrapolation. However, the extrapolations are statistically dominated, and the mild lattice-spacing dependence observed in Fig.~\ref{fig:theory-from-dm} indicates that this uncertainty is expected to be small compared with the statistical uncertainty. We therefore do not quote a separate systematic uncertainty associated with the continuum extrapolation.

\subsection{Determination of the theory scales from decay constants}
In this section we determine the theory scales 
using the kaon decay constant $f_K$ and the decay constant of unmixed pseudo-scalar
$s\bar s$ meson $f_{\eta_s}$. The decay constant $f_{\eta_s}$ cannot be determined experimentally as there
is no unmixed pseudo-scalar $s \bar s$ meson, but the corresponding decay constant can be related to pion
and kaon decay constants using chiral perturbation theory \cite{Davies:2009tsa}. 
We will use the values of $a f_K$ and $af_{\eta_s}$ obtained by HotQCD Collaboration 
in 2+1 flavor QCD \cite{HotQCD:2014kol}. The values of $f_K$ and $f_{\eta_s}$
are very sensitive to the strange quark mass. Unfortunately, the value of the strange quark mass in Ref. \cite{HotQCD:2014kol} does not follow the lines of constant physics well for $\beta>6.88$. Therefore, here we will only use 
the pseudo-scalar meson decay constant for 6.515 $ \le \beta \le $ 6.880.

Using  results of Ref. \cite{HotQCD:2014kol} we form the dimensionless products 
$f_K w_0, f_K\sqrt{t_0}, f_{\eta_s} w_0$, $f_{\eta_s}\sqrt{t_0}$, and study their dependence on the lattice spacings. In general, we find that the lattice-spacing dependence of these combinations is significantly reduced when the $t_0$ and $w_0$ scales are extracted from the improved gauge-action density.  Therefore, in the following analysis we rely
exclusively on the gradient flow scales obtained using the improved discretization. We perform continuum extrapolation of these products in $\alpha_b a^2/w_0^2$. Sample continuum extrapolations are shown in Fig. \ref{fig:decay-flow}. 
To estimate the systematic errors of the continuum extrapolation we perform
continuum fits with and without the data from the coarsest lattice corresponding
to $\beta=6.515$. The difference between these two fits is used as an estimate 
of the systematic uncertainty. Based on the previous lattice calculations \cite{Davies:2009tsa} we know that the above quantities are not very sensitive to the light quark mass. Therefore, we expect that using results obtained for
$m_l=m_s/20$ instead of the physical value of $m_l=m_s/27$ should not effect the determination of the theory scale at present level of accuracy.
\begin{figure*}[tbh]
\centerline{
\includegraphics[width=0.5\textwidth]{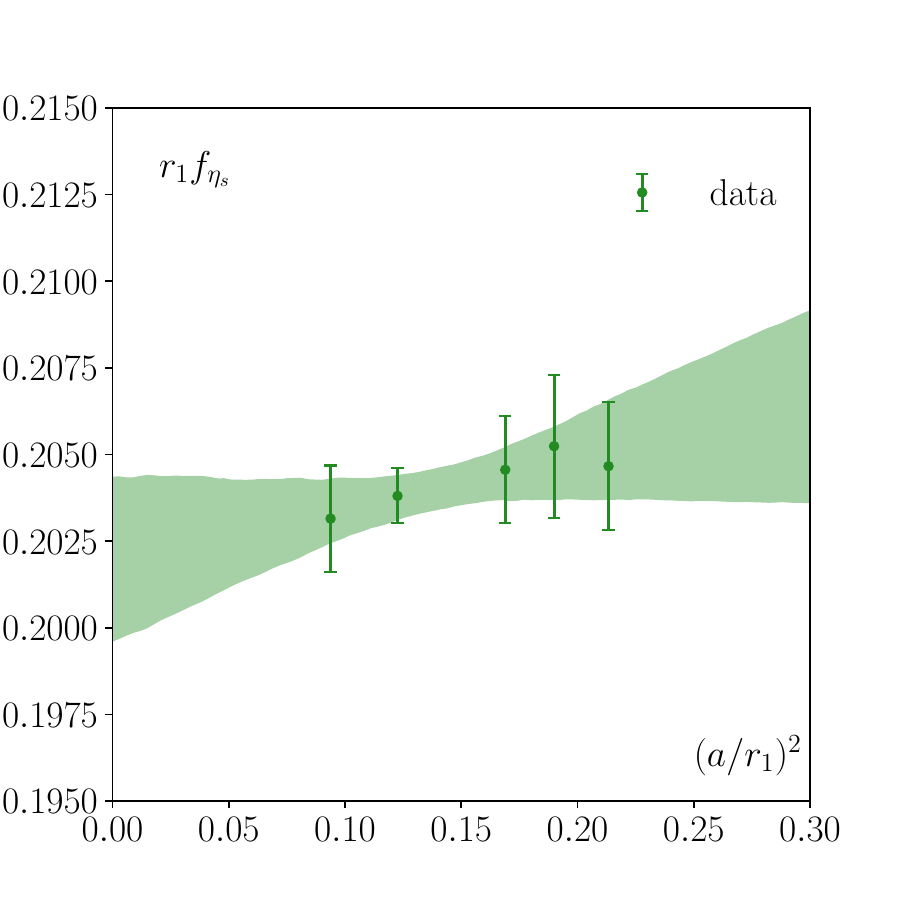}
\includegraphics[width=0.5\textwidth]{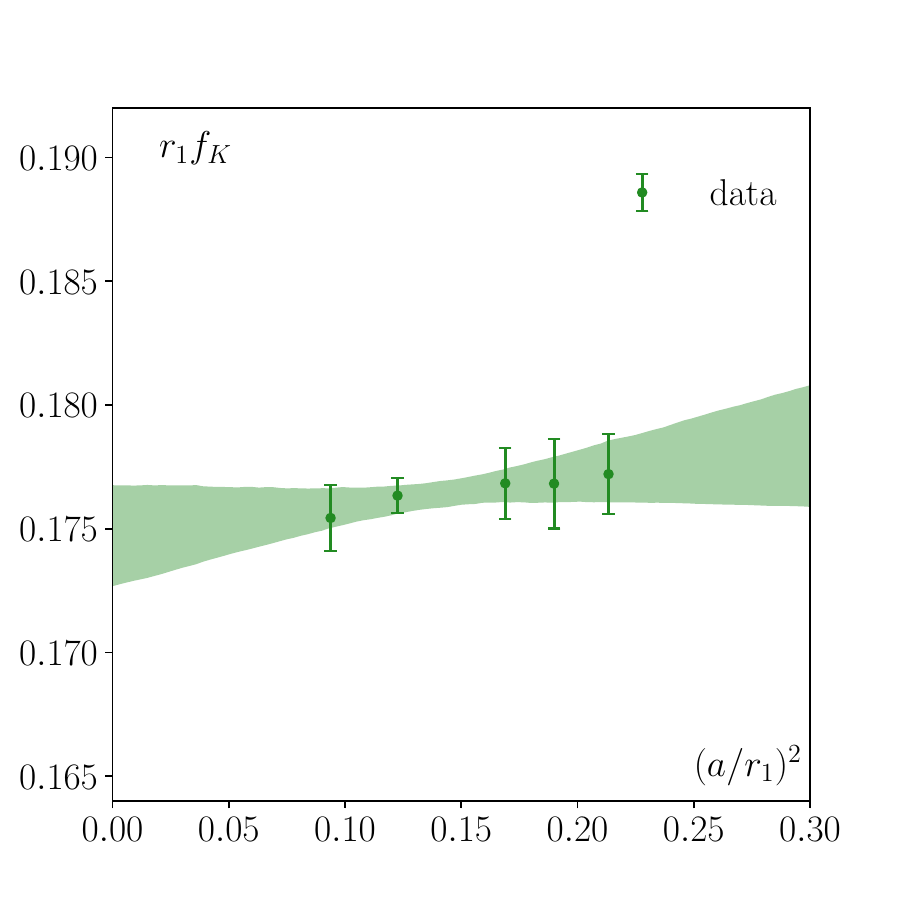}
}
\caption{The lattice spacing dependence and continuum extrapolation of $r_1f_{\eta_s}$ (left) and $r_1f_K$ (right).
}
\label{fig:decay-r1-revisit}
\end{figure*}

\begin{figure*}[tbh]
\centerline{
\includegraphics[width=0.33\textwidth]{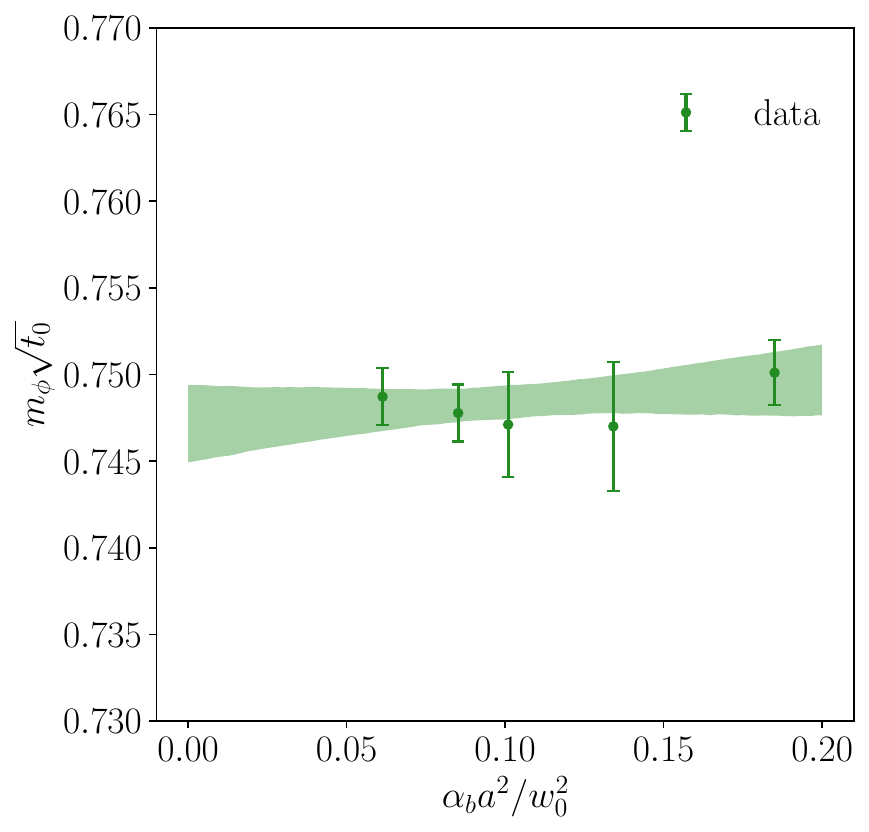}
\includegraphics[width=0.33\textwidth]{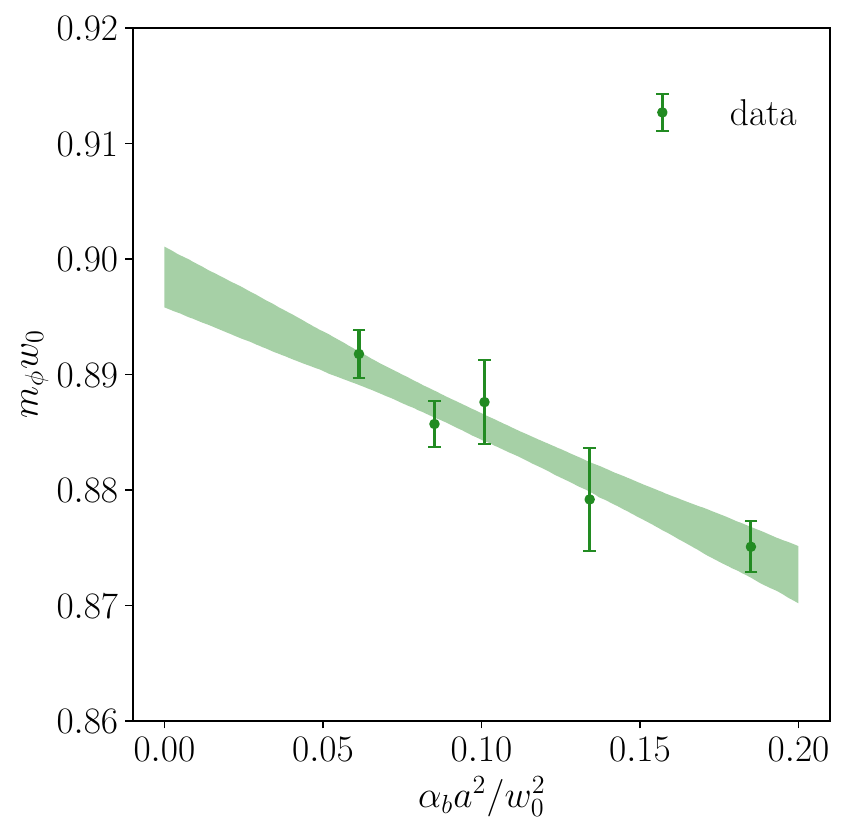}
\includegraphics[width=0.33\textwidth]{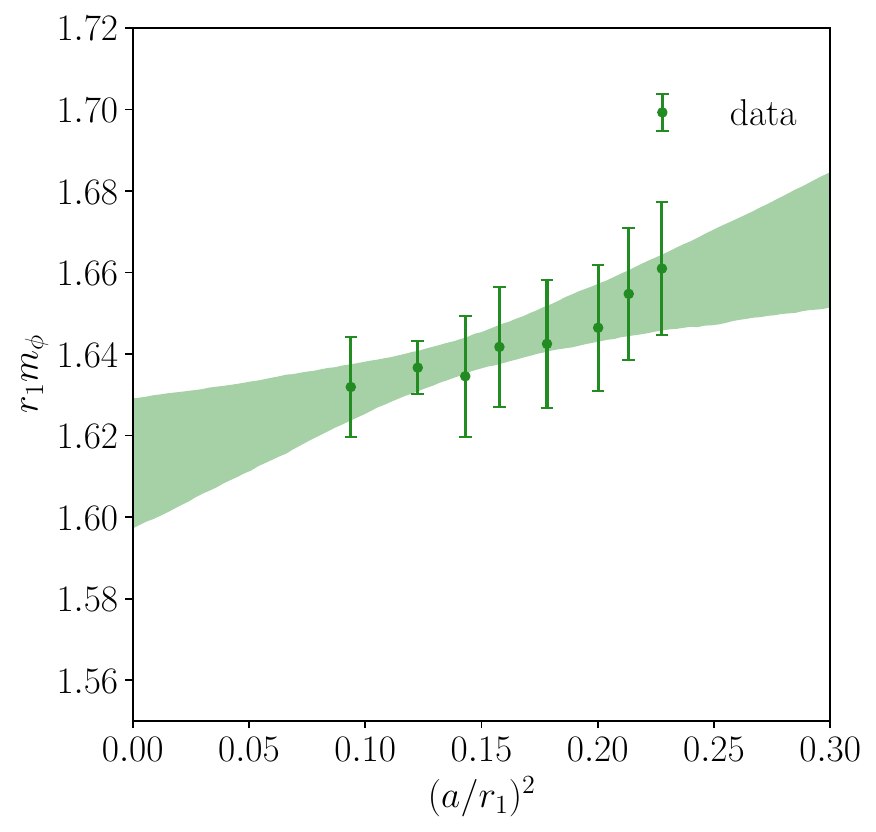}
}
\caption{
Continuum extrapolation of $m_\phi\sqrt{t_0}$ (left), $m_\phi w_0$ (middle) and $m_\phi r_1$ (right). 
}
\label{fig:phimass-flow}
\end{figure*}

Using the continuum‐extrapolated results for the above products together with the values $f_{\eta_s} = 128.34(85),\text{MeV}$ \cite{Davies:2009tsa} and $f_K = 155.7(7)/\sqrt{2},\text{MeV}$ \cite{FlavourLatticeAveragingGroupFLAG:2024oxs}, we obtain the values of the $w_0$ and $t_0$ scales in physical units. These results are presented in Tab.~\ref{tab:flow-from-decay-womass}, where we indicate 
the different sources of errors. 
By combining the statistical uncertainty, the  systematic uncertainties,
and the uncertainty in the input scale in quadrature, we obtain at the final values for the gradient flow scales determined from the meson decay constants:
\begin{equation}
\begin{split}
\sqrt{t_0}&=0.14354\ ( 99 ) ~{\rm fm}, ~{\rm via}~f_{\eta_s},\\
\sqrt{t_0}&=0.14443\ (83) ~{\rm fm}, ~{\rm via}~f_K,\\
w_0&=0.17315\ ( 129)~{\rm fm}, ~{\rm via}~f_{\eta_s},\\
w_0&=0.17425\ (89)~ {\rm fm}, ~{\rm via}~f_K.
\end{split}
\label{eq:w0-from-decay}
\end{equation}
We see that the values of the gradient flow scales determined from $f_{\eta_s}$ and $f_K$ are consistent within errors. The above results on the gradient flow 
scales based on the pesudo-scalar meson decay constants
are consistent with those determined via bottomonium mass splittings Eq~(\ref{eq:theory-DM}) and 
the FLAG average \cite{FlavourLatticeAveragingGroupFLAG:2024oxs}.

Similarly, we determine the $r_1$ scale from the decay constants.
For this we use the $r_1/a$ values determined in Ref. \cite{HotQCD:2014kol}
for $\beta = 6.460,\ 6.515,\ 6.575,\ 6.740,$ and $6.88$. 
We note that here we have also  used the results from $\beta = 6.460$ lattice from Ref. \cite{HotQCD:2014kol}, that was not used in the determination of the gradient flow scales listed in Tab.~\ref{tab:param-zeroT}.
We perform the continuum
extrapolations of the dimensionless products $r_1 f_K$ and $r_1 f_{\eta_s}$ 
assuming linear dependence in  $\alpha_b a^2/w_0^2$.
The continuum extrapolation of $r_1f_{\eta_s}$ and $r_1f_K$ are shown in Fig.~\ref{fig:decay-r1-revisit}. To estimate the systematic uncertainty of
the continuum extrapolation we perform extrapolation with and without the data
with the largest lattice spacing corresponding to $\beta=6.460$, and use
the corresponding difference as the systematic error.
The values for $r_1$ together with different sources of uncertainties
are summarized in Tab. \ref{tab:flow-from-decay-womass}. Combining
the different uncertainties in quadrature we obtain: 
\begin{equation}
\begin{split}
r_1&=0.3105\ ( 43 ) ~{\rm fm}, ~{\rm via}~f_{\eta_s},\\
r_1&=0.3131\ (40) ~{\rm fm}, ~{\rm via}~f_K.
\end{split}
\label{eq:r1-from-decay}
\end{equation}
The values of $r_1$ obtained from $f_K$ and $f_{\eta_s}$ agree with each other as well
with the value of $r_1$ obtained from the bottomonium splitting.

\subsection{Scale determination through the $\phi$ meson mass}

Finally we determine the theory scales using the $\phi$-meson mass, $m_\phi$. 
Using $m_\phi$ to set the scale may not sound to be ideal since the $\phi$ meson is a resonance. 
However, its width is only about 4 MeV, i.e. 0.4\% of its
mass. Therefore, we may expect that the errors of not treating the $\phi$
meson as a resonance should result in $0.4\%$ error or less, which is
smaller than the other sources of errors in our analysis.
Using the values of the $\phi$ meson mass from Ref. \cite{HotQCD:2014kol} we performed continuum extrapolation for the products $\sqrt{t_0} m_\phi$ and $w_0 m_\phi$ using the $\beta$=6.423, 6.55, 6.664, 6.740 and 6.88 lattices.
We fit the above dimensionless combinations with $\alpha_b a^2/w_0^2$-form.
While the lattice results seem to follow the $\alpha_b a^2$ trend, the $\chi^2$
of the fit appears to be large. We suspect that this is  most likely due to the 
fact that the errors on the $\phi$ meson mass quoted in Ref. \cite{HotQCD:2014kol}
are underestimated. Therefore, we rescale the errors on the $\phi$ meson
mass by a factor of 2.5 before performing the $\alpha_b a^2/w_0^2$ extrapolations. Then
the continuum extrapolations of $\sqrt{t_0} m_\phi$ and $w_0 m_\phi$ have 
$\chi^2/\text{d.o.f.}$ around 1. 
The corresponding continuum extrapolations are shown in Fig.~\ref{fig:phimass-flow}.
To estimate the systematic uncertainties of
the continuum extrapolation we perform extrapolations with both including
and excluding the data for the largest lattice spacing corresponding 
to $\beta=6.423$, and treating the corresponding difference as the
systematic error. Finally we need to consider the error from the input scale. 
We will use the width of the $\phi$ meson of $4.25$ MeV \cite{ParticleDataGroup:2024cfk} as the error on the input scale
instead of the error of its mass for the reasons explained above.
The values of $t_0$ and $w_0$ in physical units obtained from
the $\phi$ meson mass are shown in Tab.~\ref{tab:flow-from-decay-womass},
where we also indicate various sources of errors. 
Lattice calculations indicate that the $\phi$ meson mass is not very sensitive to the light sea quakr mass \cite{Aubin:2004wf}. Therefore,
using the HISQ lattice results on the $\phi$ meson mass obtained for $m_l=m_s/20$ instead of $m_l=m_s/27$ should not affect the determination of the theory scales.

To obtain the $r_1$ scale we consider 
product $r_1 m_\phi$ using the lattice results for $\beta$=6.423, 6.460, 6.488, 6.550, 6.608, 6.664, 6.740 and 6.88. We again perform the continuum extrapolations
using the $\alpha_b a^2/w_0^2$ form of $r_1 m_\phi$. The $\chi^2/\text{d.o.f.}$ of the fits is close to one. To estimate the systematic errors of the continuum extrapolations we have performed fits with and without the coarsest lattice corresponding
to $\beta=6.423$. The value of $r_1$ in physical units obtained from
the $\phi$ meson mass is shown in Tab.~\ref{tab:flow-from-decay-womass},
including the various sources of errors.

Using the results for $t_0$, $w_0$ and $r_1$ from Tab.~\ref{tab:flow-from-decay-womass} and combining the different sources of uncertainties  as in the previous subsection, we arrive at the final values of the theory scales determined from the $\phi$-meson mass
\begin{equation}
\begin{split}
\sqrt{t_0}&=0.14461\ ( 80 ) ~{\rm fm},~{\rm via}~m_{\phi},\\
w_0&=0.17389\ (78)~ {\rm fm}, ~{\rm via}~m_\phi,\\
r_1& =0.3122\ (33)\ \mathrm{fm,\ via\ } m_{\phi},
\end{split}
\label{eq:scales-from-mphi}
\end{equation}
which agree well with other extractions.

\begin{figure*}[tbh]
\centerline{
\includegraphics[width=0.5\textwidth]{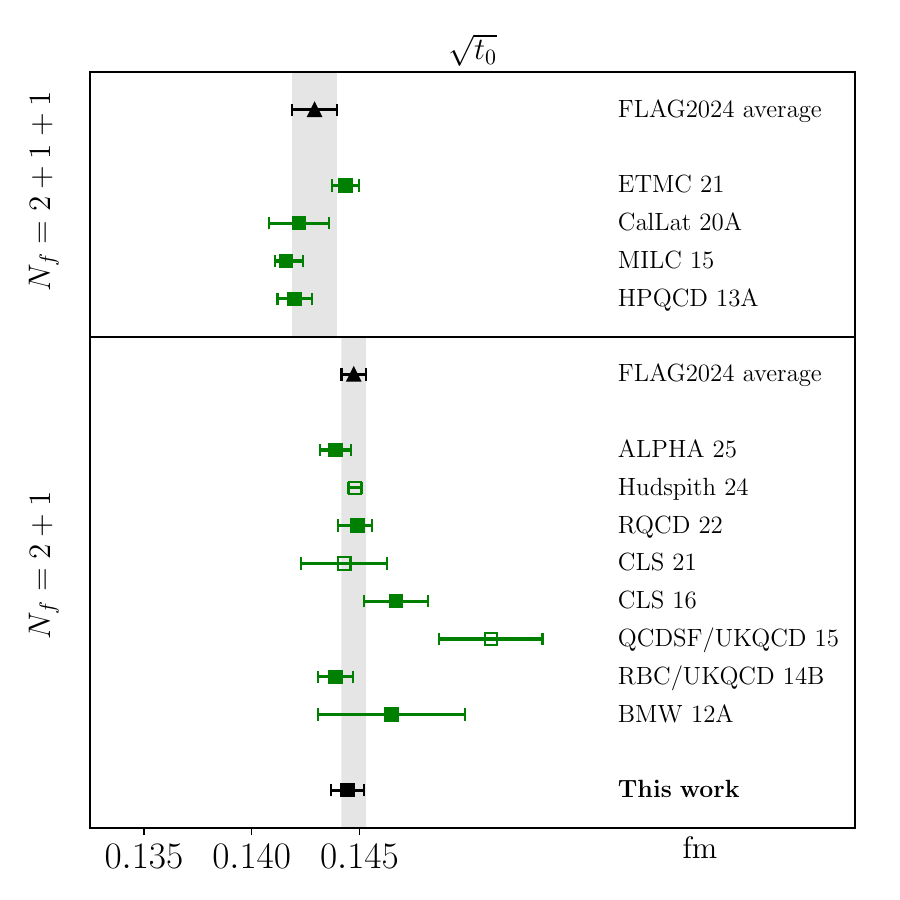}
\includegraphics[width=0.5\textwidth]{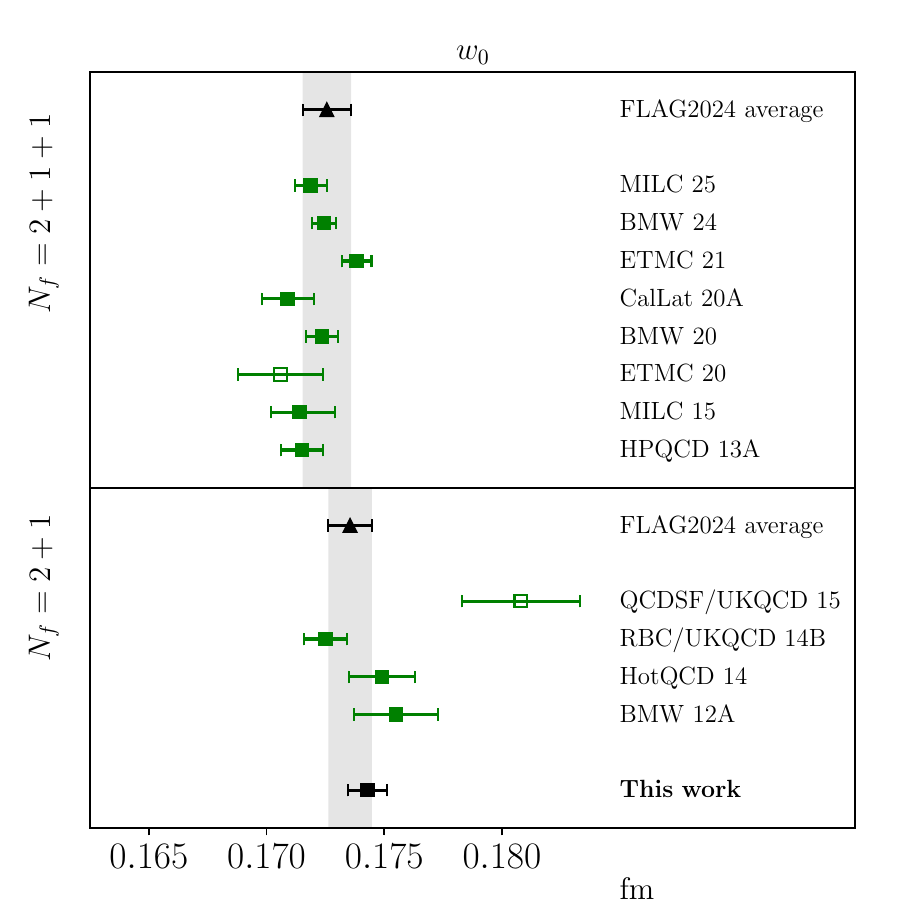}
}
\caption{A comparison of the gradient-flow scales $\sqrt{t_0}$ (left) and $w_0$ (right) determined in this work with results from the literature included in FLAG 2024 review ~\cite{FlavourLatticeAveragingGroupFLAG:2024oxs}, as well as several recent determinations~\cite{Boccaletti:2024guq,Bazavov:2025mao,Bussone:2025wlf}. 
The literature results for $\sqrt{t_0}$ shown here include:
ETMC~21~\cite{ExtendedTwistedMass:2021qui},
CalLat~20A~\cite{Miller:2020evg},
MILC~15~\cite{MILC:2015tqx},
HPQCD~13A~\cite{Dowdall:2013rya},
ALPHA~25~\cite{Bussone:2025wlf},
Hudspith~24~\cite{Hudspith:2024kzk},
RQCD~22~\cite{RQCD:2022xux},
CLS~21~\cite{Strassberger:2021tsu},
CLS~16~\cite{Bruno:2016plf},
QCDSF/UKQCD~15~\cite{Bornyakov:2015eaa},
RBC/UKQCD~14B~\cite{RBC:2014ntl},
and BMW~12A~\cite{BMW:2012hcm}.
The literature results for $w_0$ shown here include:
MILC~25~\cite{Bazavov:2025mao},
BMW~24~\cite{Boccaletti:2024guq},
ETMC~21~\cite{ExtendedTwistedMass:2021qui},
CalLat~20A~\cite{Miller:2020evg},
BMW~20~\cite{Borsanyi:2020mff},
ETMC~20~\cite{ExtendedTwistedMass:2020tvp},
MILC~15~\cite{MILC:2015tqx},
HPQCD~13A~\cite{Dowdall:2013rya},
QCDSF/UKQCD~15~\cite{Bornyakov:2015eaa},
RBC/UKQCD~14B~\cite{RBC:2014ntl},
HotQCD~14~\cite{HotQCD:2014kol},
and BMW~12A~\cite{BMW:2012hcm}.
}
\label{fig:compare-flowscales}
\end{figure*}

\begin{figure*}[tbh]
\centerline{
\includegraphics[width=0.5\textwidth]{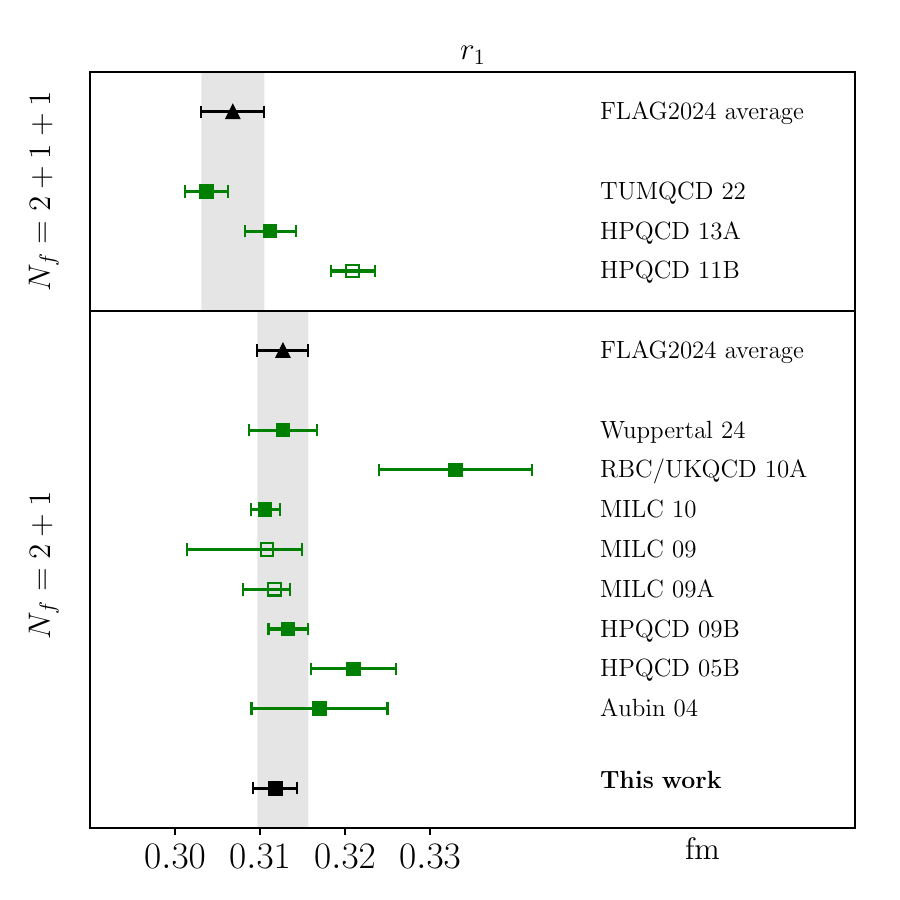}
}
\caption{A comparison of the $r_1$ scales determined in this work with those included in the FLAG~2024 review~\cite{FlavourLatticeAveragingGroupFLAG:2024oxs}, together with the recent determination of Ref.~\cite{Asmussen:2024hfw}. The literature values shown include results from
TUMQCD~22~\cite{Brambilla:2022het},
HPQCD~13A~\cite{Dowdall:2013rya},
HPQCD~11B~\cite{HPQCD:2011qwj},
Wuppertal~24~\cite{Asmussen:2024hfw},
RBC/UKQCD~10A~\cite{RBC:2010qam},
MILC~10~\cite{MILC:2010hzw},
MILC~09~\cite{MILC:2009mpl},
MILC~09A~\cite{MILC:2009ltw},
HPQCD~09B~\cite{Davies:2009tsa},
HPQCD~05B~\cite{Gray:2005ur},
and Aubin~04~\cite{Aubin:2004wf}.
}
\label{fig:compare-r1}
\end{figure*}

\subsection{Final values of the theory scales}

The determination of the theory scale from $f_K$ and $f_{\eta_s}$
are strongly correlated. The scale determinations through the $\phi$ meson
mass are dominated by systematic uncertainties. Therefore, to obtain
the final value of the theory scales 
we perform weighted average over the determinations from bottomonium mass splittings Eq.~(\ref{eq:theory-DM}) and the kaon decay constant \(f_K\) Eqs.~(\ref{eq:w0-from-decay}), (\ref{eq:r1-from-decay}). These determinations use different lattice ensembles and other sources of
errors are also uncorrelated.
Our final estimates for the theory scales are:
\begin{equation}
\begin{split}
\sqrt{t_0} &= 0.14446\ (76)\ \mathrm{fm},\\
w_0 &= 0.17429\ (83)\ \mathrm{fm}.\\
r_1 &= 0.3118\ (26)\ \mathrm{fm}.
\end{split}
\label{eq:theory-final}
\end{equation}
The determinations of the theory scale sfrom \(f_{\eta_s}\) and \(m_\phi\) are only used as cross-checks.  Reassuringly, we find that the determinations agree well with our final estimates of the scales within the quoted uncertainties.
In Fig. \ref{fig:compare-flowscales} we compare our results on the gradient flow scales with the lattice results 
reviewed by FLAG 2024 report \cite{FlavourLatticeAveragingGroupFLAG:2024oxs} both in 2+1 flavor QCD and in 2+1+1 favor QCD
as well as with two recent determinations by BMW collaboration \cite{Boccaletti:2024guq} (BMW 24) and Fermilab-MILC collaboration \cite{Bazavov:2025mao} (MILC 25) in 2+1+1 flavor QCD, 
and a very recent determination in 2+1 flavor QCD by ALPHA collaboration \cite{Bussone:2025wlf} (ALPHA 25). Our results for the gradient flow scales agree with the 2+1 flavor FLAG 2024 average. Moreover, both the FLAG average of $t_0$ and the recent determination by the ALPHA collaboration indicate that the value of $t_0$ differs between 2+1 flavor QCD and 2+1+1 flavor QCD. Our results confirm this picture, see Fig.~\ref{fig:compare-flowscales} (left). The difference in the $w_0$ values between 2+1 and 2+1+1 flavor QCD is less pronounced in the FLAG 2024 report. However, our value of $w_0$ is larger than the recent 2+1+1 flavor determinations \cite{Boccaletti:2024guq,Bazavov:2025mao}, as shown in Fig.~\ref{fig:compare-flowscales} (right). This suggests that $w_0$, like $t_0$, is also different in 2+1 flavor QCD and 2+1+1 flavor QCD. Our final result on the $r_1$ scale is compared with other results from the literature in Fig.~\ref{fig:compare-r1}, including the recent value from Ref.~\cite{Asmussen:2024hfw} (Wuppertal 24). Our result agrees with most other lattice determinations in 2+1 flavor QCD, including that of Ref.~\cite{Asmussen:2024hfw}, as shown in Fig.~\ref{fig:compare-r1}. It is in excellent agreement with the MILC collaboration’s value, $r_1 = 0.3106(18) \ \text{fm}$ \cite{MILC:2010hzw}, which was used to set the lattice spacing in the HotQCD studies. On the other hand, it is larger than the recent 2+1+1 flavor determination by the TUMQCD collaboration \cite{Brambilla:2022het}. This may suggest that, similar to the gradient flow scales, the $r_1$ scale is influenced by the presence of the dynamical charm quark. To further investigate this, we use our continuum-extrapolated value of $r_1/\sqrt{t_0}$ from Eq.~(\ref{eq:r1flow}) together with the recent determination $\sqrt{t_0} = 0.1442(6)(4)$ fm from $f_{\pi K}$ \cite{Bussone:2025wlf}. This yields $r_1 = 0.3111(20)$ fm for 2+1 flavor QCD, in good agreement with our final result. Likewise, using the continuum-extrapolated value of $r_1/w_0$ in 2+1+1 flavor QCD from Eq.~(\ref{eq:r1flow_4f}) and the recent determination $w_0 = 0.17187(96)$ fm obtained by the Fermilab–MILC collaboration from the $\Omega$-baryon mass \cite{Bazavov:2025mao}, we find $r_1 = 0.3044(16)$ fm, which agrees very well with the result from TUMQCD
$r_1=0.3037(25)$ \cite{Brambilla:2022het}.
These considerations further support the conclusion that the values of $r_1$ differ between 2+1 flavor QCD and 2+1+1 flavor QCD.

\section{Gradient flow coupling}
\label{LambdaMSbar}

\begin{figure*}[tbh]
\centerline{
\includegraphics[width=0.5\textwidth]{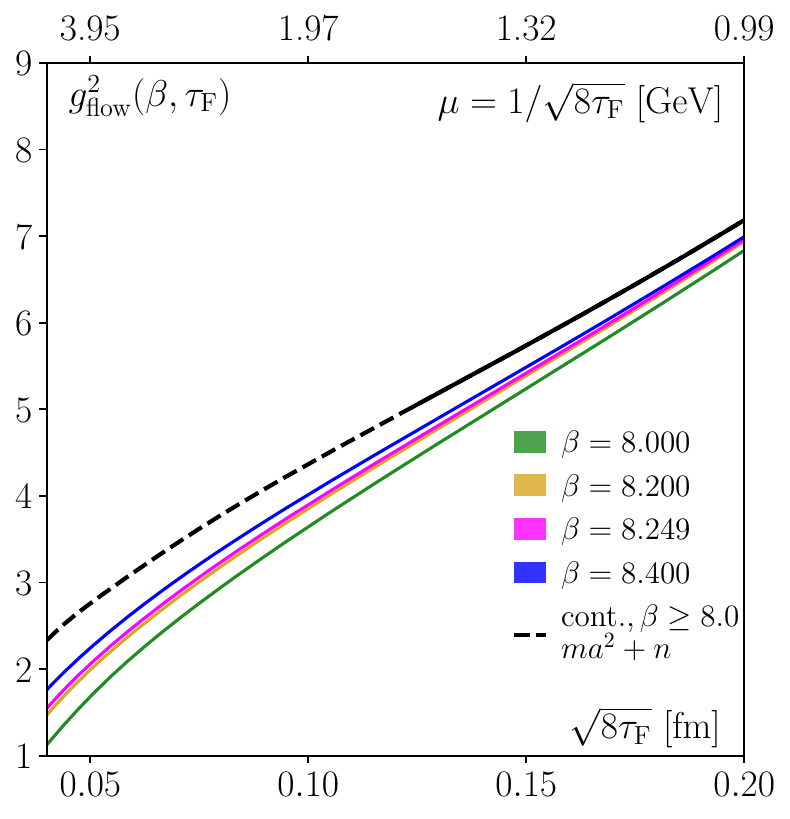}
\includegraphics[width=0.5\textwidth]{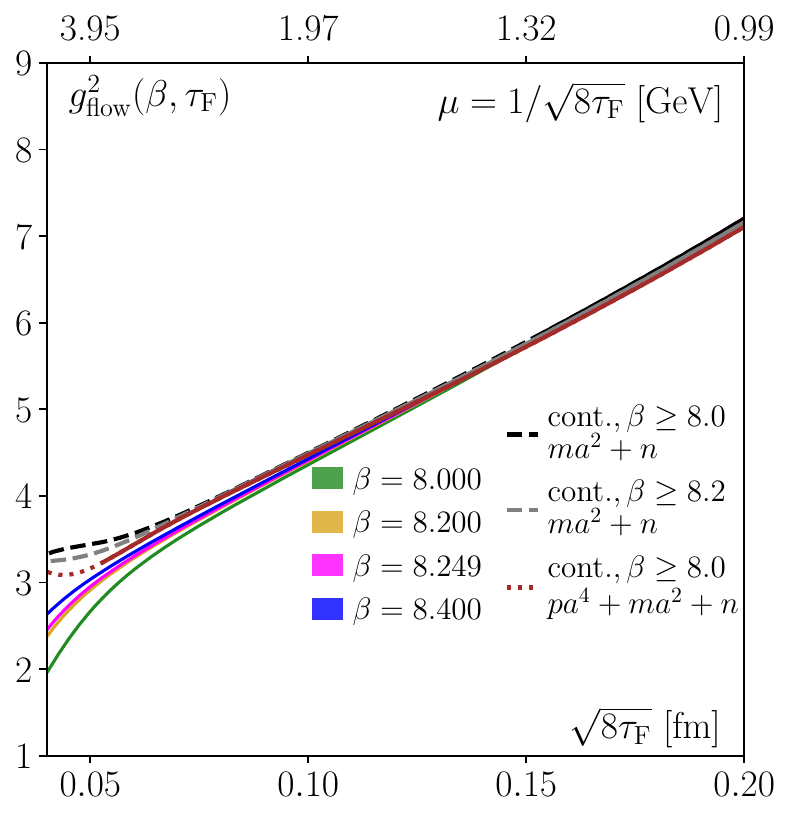}
}
\caption{
Left: the strong coupling in the gradient flow scheme measured using the clover discretization at finite lattice spacings (colored curves) and the continuum extrapolation of it using an Ansatz linear in $a^2$ (black curve). The dashed line type indicates that at the corresponding flow times the $\chi^2/$d.o.f. of the continuum extrapolation is larger than 1.5.
Right: same as the left but for the improved discretization. We also consider another Ansatz including a term quartic in $a$, which can describe the data in a much wider flow time range and try continuum extrapolation using different lattices. In these figures, the statistical errors are too small so we do not show them.
}
\label{fig:cont-extrap-summary}
\end{figure*}

The action density in Eq. (\ref{eq:action-density}) that is used to define the gradient flow scales, $t_i$ and $w_i$, can be calculated in perturbation theory 
when the flow time is sufficiently small. 
Therefore, the action density calculated with gradient flow can be used to define the gauge coupling in the gradient flow scheme (see e.g. \cite{Fodor:2012td,Hasenfratz:2019hpg})
\begin{align}
\label{eq:g2flow}
    g_{\mathrm{flow}}^2=\frac{128\pi^2}{3(N_c^2-1)}\langle \tauf^2 E\rangle,
\end{align}
where $N_c=3$ for SU(3).
This definition follows the same logic as the definition of the strong coupling \del{constant} from the static quark-antiquark potential or the corresponding
force, see e.g. Ref. \cite{Necco:2001gh}. The gradient flow coupling is defined at some energy scale, $\mu^{}_\mathrm{flow}$ that is proportional to $1/\sqrt{\tauf}$. Usually the definition $\mu^{}_\mathrm{flow}=1/\sqrt{8\tauf}$ is used. While the gradient flow coupling in Eq. (\ref{eq:g2flow}) is defined for any $\tauf$, only for sufficiently small $\tauf$ it can be related to other definitions of the coupling constant, e.g. to the conventionally used $\overline{\mathrm{MS}}$ coupling. The relation between the gradient-flow coupling and the \(\overline{\mathrm{MS}}\) coupling can be written, for a general matching scale \(\mu_{\overline{\mathrm{MS}}}=s\,\mu_{\rm flow}\), as
\begin{equation}
\begin{split}
&\alpha_f
=
\alpha_s
\bigg{[}
1+\left(k_1+B_0L_s\right)\alpha_s+
\Big{(}
k_2
+\left(
B_1
+2k_1B_0
\right)L_s+B_0^2L_s^2
\Big{)}
\alpha_s^2 \\
& 
+\Big{[}
k_3+
\left(
B_2+2k_1B_1+3k_2B_0
\right)L_s
+
\left(
\frac{5}{2}B_0B_1+3k_1B_0^2
\right)L_s^2
+B_0^3L_s^3
\Big{]}\alpha_s^3
+\cdots
\bigg{]},
\end{split}
    \label{eq:gflow}
\end{equation}
where $\alpha_f=\alpha_s^{\rm flow}(\mu_{\rm flow})=g^2_{\rm flow}/(4 \pi)$, $\alpha_s=\alpha_s^{\overline{\rm MS}}(\mu_{\overline{\rm MS}})$, $L_s \equiv \ln s^2$, $B_0=\frac{\beta_0}{4\pi}$, $\beta_0=11-\frac{2}{3}N_f$, $B_1=\frac{\beta_1}{16\pi^2}$, $\beta_1=102-\frac{38}{3}N_f$,  $B_2=\frac{\beta_2}{64\pi^3}$ and $\beta_2=\frac{2857}{2}-\frac{5033}{18}N_f+\frac{325}{54}N_f^2$ with $N_f=3$. For the special choice \(s=1\), i.e. \(\mu_{\overline{\mathrm{MS}}}=\mu_{\rm flow}\), the logarithmic terms vanish and the above expression reduces to
\begin{equation}
    \alpha_f  \equiv \frac{g^2_\mathrm{flow}}{4 \pi }= \alpha_s
    \left( 1 + k_1 \alpha_s + k_2 \alpha_s^2 +k_3 \alpha_s^3+\dots\right).
    \label{eq:gflow2}
\end{equation} 
The coefficients $k_1$ and $k_2$ have been calculated by Harlander and Neumann  \cite{Harlander:2016vzb}
\begin{equation}
\begin{split}    
    k_1 & = 1.098 + 0.008 N_f,\\
    k_2 & = -0.982 - 0.070 N_f + 0.002 N_f^2,
\end{split}
\label{eq:convert-NNLO}
\end{equation}
while $k_3$ at present is unknown. 

\begin{figure*}[tbh]
\centerline{
\includegraphics[width=0.51\textwidth]{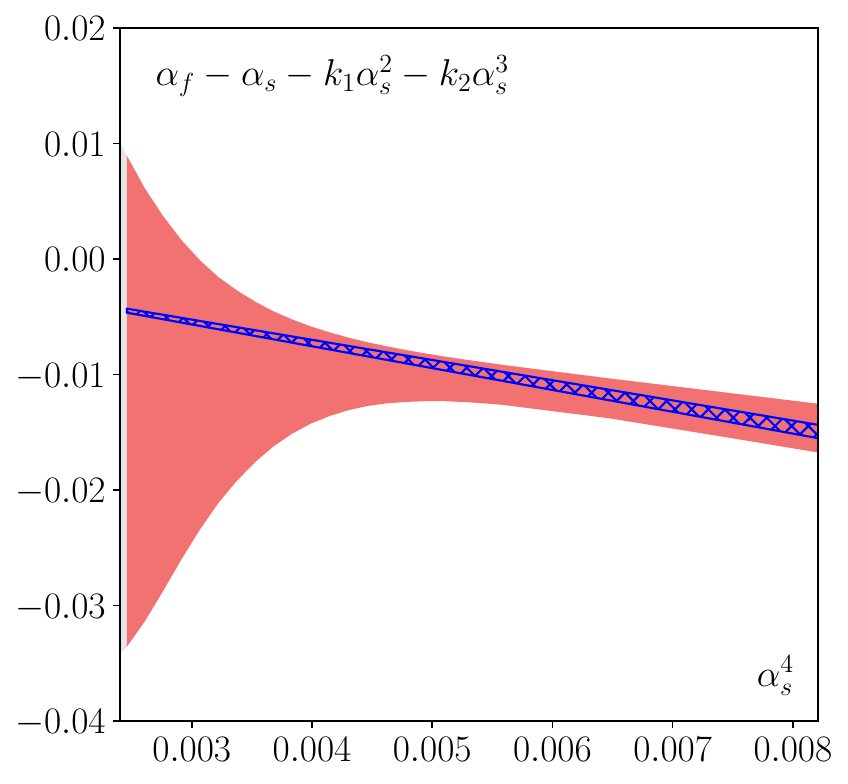}
\includegraphics[width=0.49\textwidth]{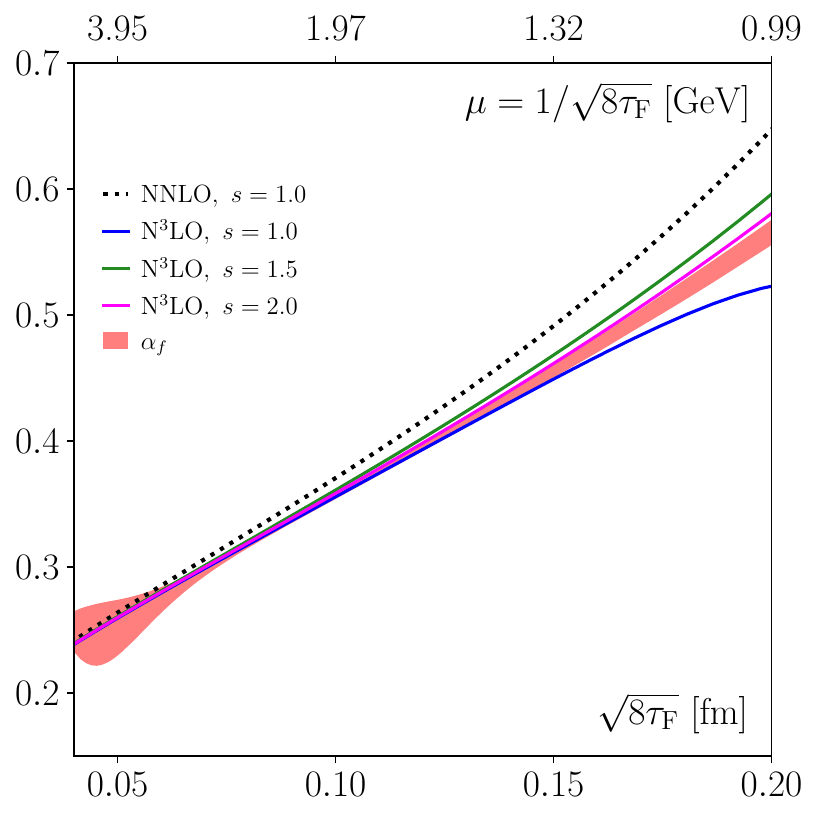}
}
\caption{
Left: $\alpha_f-\alpha_{\mathrm{s}}-k_1\alpha_{\mathrm{s}}^2-k_2 \alpha_{\mathrm{s}}^3$ as a function of $\alpha_\mathrm{s}^4$ estimated at 
scale $\mu^{\overline{MS}}=\mu_{\rm flow}$ using 4-loop running with $\Lambda^{N_f=3}_{\overline{\mathrm{MS}}}=338$ MeV. We also show the fits of $\alpha_f-\alpha_{\mathrm{s}}-k_1\alpha_{\mathrm{s}}^2-k_2 \alpha_{\mathrm{s}}^3$  with $k_3\alpha_{\mathrm{s}}^4$, see text.
Right: the gradient flow coupling obtained $\alpha_f=g_{\rm flow}^2/(4 \pi)$
on the lattice in the continuum limit compared with the perturbative result, see text.
}
\label{fig:gflow_comp}
\end{figure*}

Our goal is to calculate the gradient flow 
coupling on the lattice in the continuum
limit at small flow time, $\sqrt{8 \tauf}<0.2$ fm and see to what extent the running
of the gradient flow coupling can be described by perturbation theory. It turns out that 
lattice artifacts in the gradient flow 
coupling are quite large for $\sqrt{8 \tauf}<0.2$ fm and continuum extrapolations
are challenging.
To minimize lattice artifacts when calculating the gradient flow
coupling we only consider the highest four $\beta$ values: 8.0,~8.2,~8.249 and 8.4. The gradient flow coupling calculated at these $\beta$ values using 
the action density with the clover discretization and the improved discretization of the field strength tensor are shown in the left and right panels of Fig.~\ref{fig:cont-extrap-summary}, respectively. 
For clover discretization of the gauge action density we performed continuum extrapolation of $g_{\mathrm{flow}}^2$ assuming $a^2$-dependence, i.e. we performed fits using $ma^2+n$ form. For $\sqrt{8 \tauf}<0.12$ fm the fits had large $\chi^2$/d.o.f., larger than 1.5.
Thus for these $\tauf$ values the continuum extrapolations are not reliable. From Fig. \ref{fig:cont-extrap-summary} (left
) we see that lattice spacing dependence of $g_{\mathrm{flow}}^2$ is significant and there is also a sizable difference between
the continuum extrapolated value of gradient flow coupling and that obtained on the finest lattice. The situation is different
for the improved discretization of the gauge action density as one can see in Fig. \ref{fig:cont-extrap-summary} (right). Namely, the dependence of the gradient
flow coupling on $\beta$ is much reduced. 
In the case of the 
improved discretization we performed different continuum extrapolations. We used simple quadratic form, $ma^2+n$ form for
continuum extrapolations and
including all $\beta\ge 8.0$ lattices, or excluding the $\beta=8.0$ lattice, or using quadratic plus quartic form, i.e. $pa^4+ma^2+n$ form for continuum extrapolations and using all
$\beta\ge 8.0$ data. The corresponding continuum extrapolations are also shown in Fig. \ref{fig:cont-extrap-summary} (right). For \(\sqrt{8\tau_{\rm F}}>0.075~{\rm fm}\), the difference among different continuum extrapolations is small. For example, at \(\sqrt{8\tau_{\rm F}}=0.076~{\rm fm}\), the relative difference between the black and brown curves is \(0.7\%\), while that between the gray and brown curves is \(0.4\%\).
Furthermore, as shown in the right panel of Fig.~\ref{fig:cont-extrap-summary}, the difference between the continuum-extrapolated result and the result obtained on the finest lattice is also much smaller for the improved discretization. At \(\sqrt{8\tau_{\rm F}}=0.076~{\rm fm}\), the difference in \(g_{\mathrm{flow}}^2\) is \(0.41\) for the clover discretization, compared with only \(0.084\) for the improved discretization. We note that, in the range of interest,
\(\sqrt{8\tau_{\rm F}}\in[0.075,0.2]~{\rm fm}\), the corresponding values of
\(\sqrt{8\tau_{\rm F}}/a\) are
\([2.19,\ 5.83]\), \([2.61,\ 6.96]\), \([2.72,\ 7.27]\), and \([3.11,\ 8.29]\)
for \(\beta=8.000\), \(8.200\), \(8.249\), and \(8.400\), respectively.

Since the continuum-extrapolated result for the gradient flow coupling was obtained from lattices with $\beta \ge 8.0$ where the light quark masses are heavier than the physical ones, the sensitivity of
the gradient flow coupling to the light quark mass
should be studied. In addition, the impact of frozen topology needs to be addressed. However, because our analysis focuses on the gradient flow coupling at relatively small flow times, the influence of frozen topology is expected to be minor. This expectation is confirmed in Appendix~\ref{app:topo-analysis}, where a comparison of the flow-time dependence of $g_{\mathrm{flow}}^2$ across different topological sectors shows that the differences remain negligible up to $\sqrt{8\tauf} = 0.2$ fm. To investigate the effect of light-quark masses on the gradient flow coupling, we examine results on a slightly coarser lattice corresponding to $\beta = 7.825$, where simulations are available for two mass ratios, $m_s/m_l = 20$ and $m_s/m_l = 5$. As demonstrated in Appendix~\ref{app:topo-analysis}, the differences in $g_{\mathrm{flow}}^2$ between these two ensembles are very small, smaller than the uncertainty of the continuum extrapolation. Thus, at the current level of precision, the dependence on the light-quark masses can be neglected.

Having obtained continuum-extrapolated results for the gradient flow coupling, and having established that the effects of the light-quark mass  and frozen topology can be neglected, we can now ask to what extent the flow-time dependence of $g_{\mathrm{flow}}^2$ can be described by perturbation theory. Using the value $\Lambda_{\overline{\mathrm{MS}}}^{N_f=3} = 338(10)\ \text{MeV}$ from FLAG 2024 \cite{FlavourLatticeAveragingGroupFLAG:2024oxs}, based on Refs.~\cite{McNeile:2010ji,Chakraborty:2014aca,DallaBrida:2022eua,Petreczky:2020tky,Ayala:2020odx,Bazavov:2019qoo,Cali:2020hrj,Bruno:2017gxd,PACS-CS:2009zxm,Maltman:2008bx}, and Eq.~(\ref{eq:gflow}) with $k_3 = 0$, we obtain the next-to-next-to leading (NNLO) perturbative prediction for the gradient flow coupling. This is shown in Fig.~\ref{fig:gflow_comp}, together with the continuum-extrapolated lattice result whose uncertainty band includes both statistical and systematic contributions added in quadrature. The systematic uncertainty, arising from the continuum extrapolation, is taken as the larger of the two deviations shown in the right panel of Fig.~\ref{fig:cont-extrap-summary}: the difference between the black and brown curves, and the difference between the gray and brown curves.
A significant discrepancy is observed between the known perturbative result and the lattice data on the gradient flow coupling for $\sqrt{8 \tau_F}>0.075$ fm. 
This discrepancy may arise from higher-order perturbative corrections, non-perturbative contributions, or both. 
To test whether the discrepancy between the lattice results  and the NNLO perturbative result can be explained by the missing next-to-next-to-next-to (N$^3$LO) correction $k_3 \alpha_s^4$ in Fig. \ref{fig:gflow_comp} (left)
we show the combination 
$\alpha_f-\alpha_{\mathrm{s}}-k_1\alpha_{\mathrm{s}}^2-k_2 \alpha_{\mathrm{s}}^3$ 
as function 
for $\alpha_{\mathrm{s}}^4$ evaluated for $\mu=\mu_{\rm flow}$ for flow time 
$0.051~{\rm fm}<\sqrt{8 \tau_F}<0.137$ fm. This combination should approach zero linearly
in $\alpha_{\mathrm{s}}^4$ if the missing N$^3$LO term is the cause of
the discrepancy between the lattice and perturbative results. 
The lattice results for $\alpha_f-\alpha_{\mathrm{s}}-k_1\alpha_{\mathrm{s}}^2-k_2 \alpha_{\mathrm{s}}^3$ are compatible with linear behavior in $\alpha_s^4$ within the errors for the above flow time window. Unfortunately, at the small flow time the errors are very large.
We also performed a linear fit of $\alpha_f-\alpha_{\mathrm{s}}-k_1\alpha_{\mathrm{s}}^2-k_2 \alpha_{\mathrm{s}}^3$ with $k_3 \alpha_{\rm s}^4$ for $0.051~{\rm fm}<\sqrt{8 \tau_F}<0.137$  fm to estimate $k_3$. This fit gives $k_3=-1.82(7)$ with $\chi^2/dof=1.3$ as is shown in Fig. \ref{fig:gflow_comp} (left) as the blue band. Restricting the fit range to  $0.051~{\rm fm}<\sqrt{8 \tau_F}<0.1$ fm gives $k_3 =-1.94(9)$ and $\chi^2/dof=1.2$, i.e. the same result within errors.
The absolute value of $k_3$ is similar to that of $k_2$ and $k_1$, thus the interpretation of the corresponding term as the missing higher order correction is reasonable.
Using this value of $k_3=-1.82$ we show the corresponding  N$^3$LO pertubative estimates  given Eq. (\ref{eq:gflow}) for $s=1,~1.5$ and 2, which are also shown in the right panel of Fig. \ref{fig:gflow_comp}. As one can see 
from the figure adding the N$^3$LO reduces the discrepancy between the lattice and the perturbative result for $\sqrt{8 \tau_F}<0.15$ fm. For larger flow time there is a large scale ($s$) dependence
and there are significant deviations from the lattice result. For these flow times perturbative
corrections beyond N$^3$LO as well as non-perturbative corrections become important and
the running of the gradient flow coupling clearly cannot be described by perturbation theory.
We also note that while the N$^3$LO perturbative estimate roughly agrees with the lattice
results its scale dependence is non-monotonic. 
The non-monotonic dependence on the renormalization scale makes the determination of the
perturbative error difficult.
Because of this fact and the relatively large value 
of $\alpha_f$ a reliable extracting of the $\Lambda$ parameter from the present flow time window
is not possible.

\section{Conclusions}
\label{summary}
In this paper we determined the gradient flow scales $t_0$, $w_0$,
$t_2$ and $w_2$
in 2+1 flavor QCD using HISQ action in a wide range of lattice spacing.
This allows us to use these scales to set the lattices spacing in the high temperature calculations performed by HotQCD collaboration. The
physical values of $t_0$ and $w_0$ have been  determined 
using bottomonium mass splitting, $f_K$, $f_{\eta_s}$ and the $\phi$ meson mass.  To
obtain the final value of the gradient flow scales we 
performed the
weighted average of the determinations of the scales using bottomonium splitting
and kaon decay constant, resulting in $\sqrt{t_0} = 0.14446(76)$~fm and $w_0 = 0.17429(83)$~fm as our final result. These values agree well with other lattice determinations in 2+1 flavor QCD, but are larger than the recent determinations in 2+1+1
flavor QCD, confirming earlier expectations that the gradient flow 
scales are affected by the dynamical charm quark. We also revisited the determination of the $r_1$ scale from the heavy quark potential.
We find $r_1 = 0.3118(26)$~fm, which agrees with other lattice QCD determinations in 2+1 flavor QCD but is larger than the recent determination of $r_1$ in 2+1+1 flavor QCD by TUMQCD collaboration
\cite{Brambilla:2022het}. This implies that also the $r_1$ scale is affected by the dynamical charm quark. Furthermore, we determined 
the ratios of the gradient flow scales and $r_1$ scale, which
supports the above assertion.

Finally, we determined the gradient flow coupling in the continuum
limit for small flow time. 
We find that the running of the gradient flow coupling could be compatible with
perturbation theory  up to $\sqrt{8\tauf}=0.15$ fm once the missing N$^3$LO correction is included. To test this assertion and possibly to determine the QCD $\Lambda$ parameter
an explicit calculation of the N$^3$LO term in perturbation theory is needed as well as lattice calculations at smaller flow time.
\section*{Acknowledgements}
This material is based upon work supported by The U.S. Department of Energy, Office of Science, Office of Nuclear Physics through Contract No.~DE-SC0012704, within the frameworks of Scientific Discovery through Advanced Computing (SciDAC) award Fundamental Nuclear Physics at the Exascale and Beyond, and with "Heavy Flavor Theory for QCD Matter (HEFTY)" topical collaboration in Nuclear Theory.
R.L. was supported by the Ministry of Culture and Science of the State of Northrhine Westphalia (MKW NRW) under the funding code NW21-024-A (NRW-FAIR). J.H.W.’s research has been funded by the Deutsche Forschungsgemeinschaft (DFG, German Research Foundation)---Projektnummer 417533893/GRK2575 Rethinking Quantum Field Theory. J.H.W. acknowledges the support by the State of Hesse within the Research Cluster ELEMENTS (Project ID 500/10.006). 

This research used awards of computer time provided by the National Energy Research Scientific Computing Center (NERSC), a U.S. Department of Energy Office of Science User Facility located at Lawrence Berkeley National Laboratory, operated under Contract No. DE-AC02- 05CH11231. Computations for this work were carried out in part on facilities of the USQCD Collaboration, which are funded by the Office of Science of the U.S. Department of Energy.

All computations in this work were performed using \texttt{SIMULATeQCD}~\cite{Mazur:2021zgi, Bollweg:2021cvl, HotQCD:2023ghu}. 

\section*{Appendix}
\label{app:Appendix}
\appendix

\section{Gradient flow scales computed on the finite lattices}
\label{app:flowscales}
In Tab.~\ref{tab:flowscales20} and \ref{tab:flowscales5} we summarize the flow scales in lattice units $\sqrt{t_i}/a$ and $w_i/a$ calculated on the ensembles listed in Tab.~\ref{tab:param-zeroT} using both clover and improved discretization.

\begin{table*}[tbh]
\centering
\tiny
\begin{tabular}{c|cccccccc}
\hline\hline
$\beta$ &
$({\sqrt{t_0}/a})_{\mathrm{clo.}}$ &
$({\sqrt{t_0}/a})_{\mathrm{imp.}}$ &
$({\sqrt{t_2}/a})_{\mathrm{clo.}}$ &
$({\sqrt{t_2}/a})_{\mathrm{imp.}}$ &
$({w_0/a})_{\mathrm{clo.}}$ &
$({w_0/a})_{\mathrm{imp.}}$ &
$({w_2/a})_{\mathrm{clo.}}$ &
$({w_2/a})_{\mathrm{imp.}}$
\\ \hline
6.423 &
- & 0.9464(2) &
- & 0.7125(1) &
- & 1.1042(4) &
- & 0.8946(3)
\\
6.515 &
- & 1.0393(2) &
- & 0.7730(1) &
- & 1.2196(5) &
- & 0.9937(4)
\\
6.550 &
- & 1.0781(2) &
- & 0.7989(1) &
- & 1.2689(6) &
- & 1.0344(4)
\\
6.575 &
- & 1.1050(3) &
- & 0.8175(1) &
- & 1.3009(7) &
- & 1.0613(5)
\\
6.664 &
- & 1.2087(3) &
- & 0.8896(2) &
- & 1.4360(3) &
- & 1.1660(6)
\\
6.740 &
1.4203(3) & 1.3045(3) &
1.1037(1) & 0.9570(1) &
1.5577(6) & 1.5451(7) &
1.2658(4) & 1.2626(4)
\\
6.880 &
1.6037(5) & 1.4975(4) &
1.2328(2) & 1.0939(2) &
1.7920(12) & 1.7835(12) &
1.4545(7) & 1.4563(7)
\\
7.030 &
1.8259(5) & 1.7298(5) &
1.3891(2) & 1.2593(2) &
2.0730(13) & 2.0674(13) &
1.6832(8) & 1.6883(8)
\\
7.150 &
2.0217(8) & 1.9334(7) &
1.5276(4) & 1.4056(3) &
2.3159(21) & 2.3121(22) &
1.8826(13) & 1.8893(14)
\\
7.280 &
2.2636(13) & 2.1831(13) &
1.6076(5) & 1.5841(4) &
2.6203(36) & 2.6178(37) &
2.1302(15) & 2.1380(16)
\\
7.373 &
2.4549(19) & 2.3799(19) &
1.8320(7) & 1.7244(7) &
2.8575(48) & 2.8558(50) &
2.3258(27) & 2.3341(27)
\\
7.596 &
2.9847(28) & 2.9218(30) &
2.2048(12) & 2.1116(12) &
3.5194(73) & 3.5193(72) &
2.8653(40) & 2.8740(44)
\\
7.825 &
3.634(9) & 3.586(7) &
2.669(3) & 2.591(3) &
4.292(24) & 4.304(19) &
3.510(13) & 3.524(11)
\\
\hline\hline
\end{tabular}
\caption{The dimensionless gradient flow scales $\sqrt{t_i}/a$ and $w_i/a$ determined from different discretization at different $\beta$ for the $m_s/m_l=20$ ensembles.}
\label{tab:flowscales20}
\end{table*}

\begin{table*}[tbh]
\centering
\tiny
\begin{tabular}{c|cccccccc}
\hline\hline
$\beta$ &
$({\sqrt{t_0}/a})_{\mathrm{clo.}}$ &
$({\sqrt{t_0}/a})_{\mathrm{imp.}}$ &
$({\sqrt{t_2}/a})_{\mathrm{clo.}}$ &
$({\sqrt{t_2}/a})_{\mathrm{imp.}}$ &
$({w_0/a})_{\mathrm{clo.}}$ &
$({w_0/a})_{\mathrm{imp.}}$ &
$({w_2/a})_{\mathrm{clo.}}$ &
$({w_2/a})_{\mathrm{imp.}}$
\\ \hline
7.030 &
1.806(3) & 1.711(3) &
1.380(1) & 1.252(1) &
2.022(5) & 2.015(5) &
1.652(4) & 1.656(4)
\\
7.825 &
3.622(9) & 3.568(8) &
2.664(4) & 2.587(3) &
4.262(19) & 4.261(18) &
3.486(14) & 3.502(13)
\\
8.000 &
4.228(33) & 4.184(33) &
3.096(12) & 3.027(11) &
4.978(83) & 4.978(83) &
4.083(47) & 4.089(47)
\\
8.200 &
5.083(79) & 5.044(78) &
3.688(21) & 3.629(20) &
6.156(345) & 6.156(345) &
4.999(188) & 5.006(189)
\\
8.249 &
5.262(50) & 5.225(50) &
3.837(26) & 3.780(25) &
6.233(81) & 6.233(81) &
5.109(60) & 5.115(60)
\\
8.400 &
6.076(108) & 6.045(106) &
4.399(32) & 4.347(29) &
7.293(330) & 7.297(333) &
5.954(199) & 5.962(201)
\\
\hline\hline
\end{tabular}
\caption{Same as Tab.~\ref{tab:flowscales20} but for the $m_s/m_l=5$ ensembles.}
\label{tab:flowscales5}
\end{table*}

\section{Influence of topological freezing and the light quark
mass on flow observables}
\label{app:topo-analysis}

\begin{figure*}[tbh]
\centerline{
\includegraphics[width=0.33\textwidth]{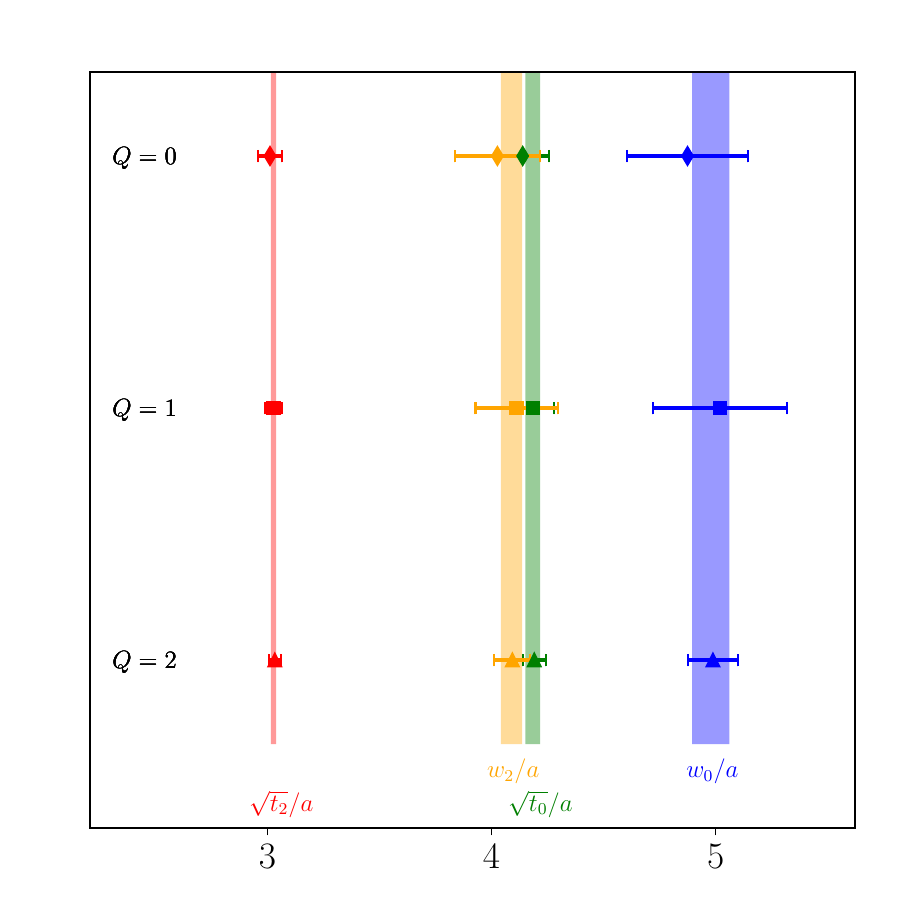}
\includegraphics[width=0.33\textwidth]{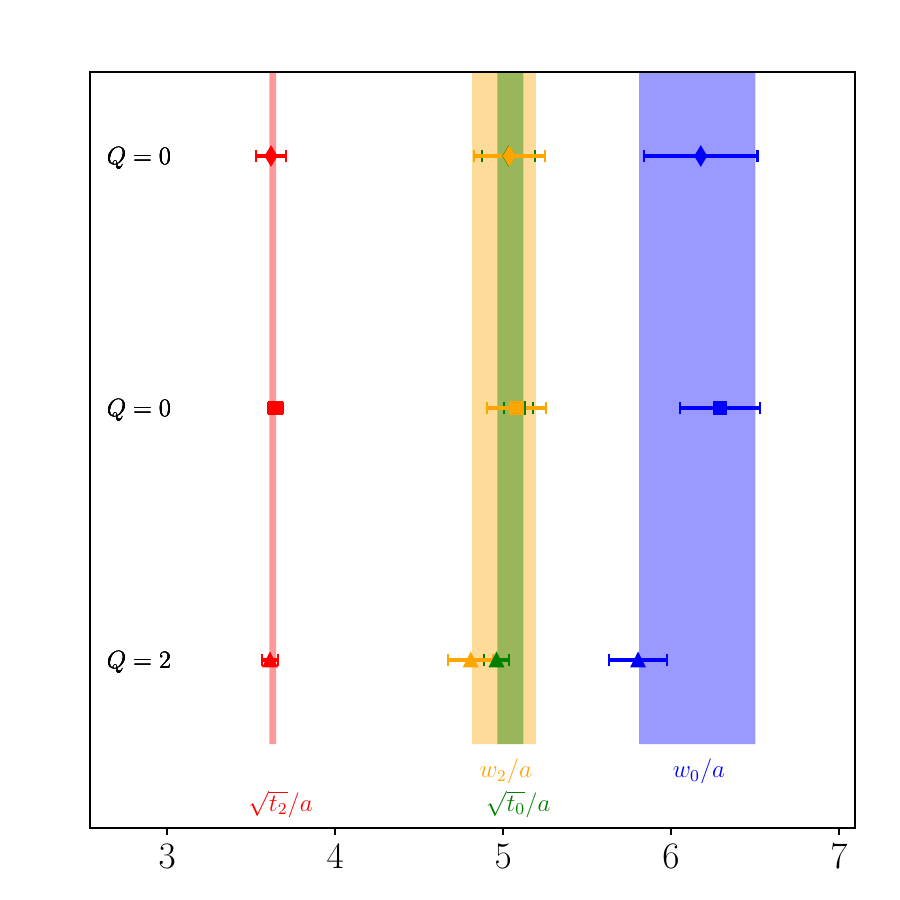}
\includegraphics[width=0.33\textwidth]{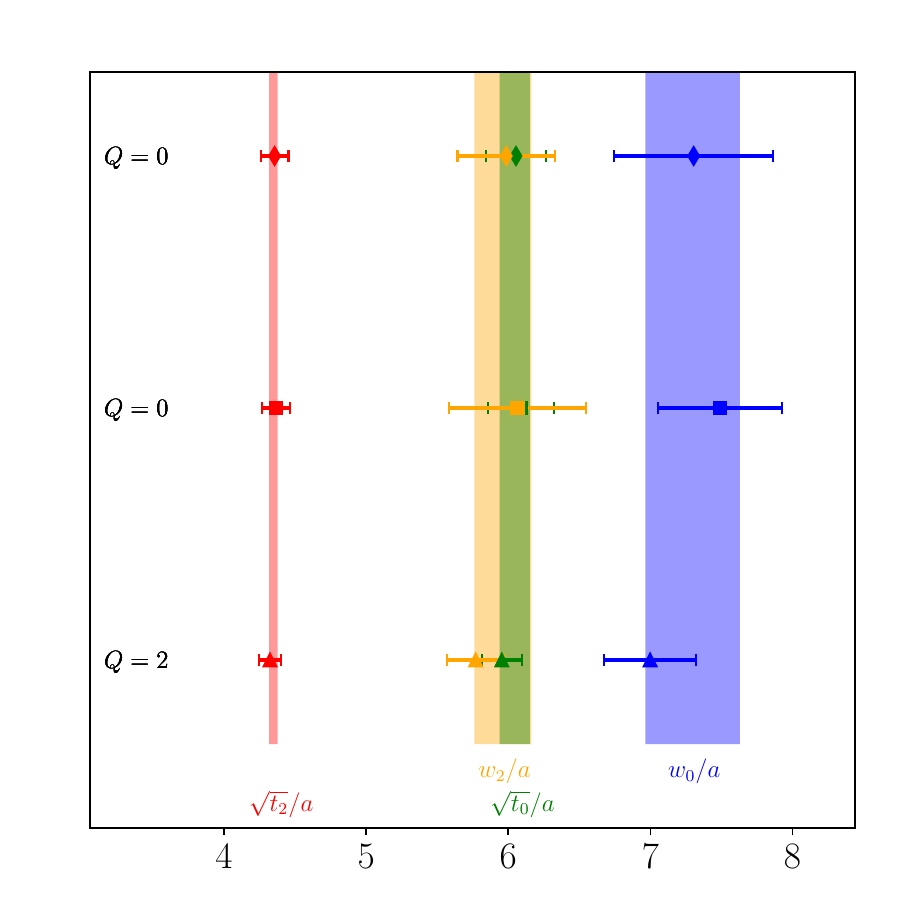}
}
\caption{The gradient flow scales in lattice units $\sqrt{t_i}/a$ and $w_i/a$ determined from the improved discretization at  $\beta=$8.0 (left), 8.2 (middle), 8.4 (right) in the different topological sectors. The mean and error for each stream are estimated from the original configurations instead of the binned ones, because the latter has only $\sim$10-30 statistics for each stream. The bands denote the average over different streams, estimated from the bootstrap samples drawn from the binned configurations.}
\label{fig:compare-topo}
\end{figure*}

For $\beta\ge 8.0$ the topological charge, $Q$ is frozen 
during Monte-Carlo
sampling. For $\beta=8.0,~8.2$ and 8.4 we have three independent Monte-Carlo streams labeled as a,b and c, some of which have differed values of $Q$.
In Tab. \ref{tab:flowscales-stream} and in Fig. \ref{fig:compare-topo}
we show the gradient flow scales for different streams. 
As one can see, within the statistical uncertainties of the individual streams, there is no statistically significant difference between the gradient-flow scales corresponding to different values of \(Q\).

\begin{table*}[tbh]
\centering
\begin{tabular}{c|c|c|c|c}
 \hline  \hline 
 $\beta$ & \multicolumn{4}{c}{8.000}   \\ \hline
  & steam a & steam b & stream c &   \\ \hline
  &  $Q=2$  &  $Q=1$  &   $Q=0$  &  average  \\ \hline 
$\sqrt{t_0}/a$(impr.) & 4.191(51) & 4.185(96) & 4.139(118) & 4.184(33) \\
$\sqrt{t_2}/a$(impr.) & 3.033(27) & 3.027(37) & 3.012(52) & 3.027(11) \\
$w_0/a$(impr.) & 4.988(110) & 5.019(299) & 4.875(269) & 4.978(83) \\
$w_2/a$(impr.) & 4.093(80) & 4.112(183) & 4.026(191) & 4.089(47) \\
\hline 
\end{tabular}

\begin{tabular}{c|c|c|c|c}
\hline 
 $\beta$ & \multicolumn{4}{c}{8.200}  \\ \hline
  & steam a & steam b & stream c &  \\ \hline
  &  $Q=2$  &  $Q=0$  &   $Q=0$  &  average \\ \hline 
$\sqrt{t_0}/a$(impr.) &  4.961(76) & 5.090(86) & 5.034(158) & 5.044(78) \\
$\sqrt{t_2}/a$(impr.) &  3.612(45) & 3.645(43) & 3.618(89) & 3.629(20) \\
$w_0/a$(impr.) & 5.804(172) & 6.291(238) & 6.177(338) & 6.156(345) \\
$w_2/a$(impr.) & 4.807(133) & 5.080(177) & 5.037(212) & 5.006(189) \\
\hline 
\end{tabular}

\begin{tabular}{c|c|c|c|c}
\hline 
 $\beta$ & \multicolumn{4}{c}{8.400}  \\ \hline
  & steam a & steam b & stream c &  \\ \hline
  &  $Q=2$  &  $Q=0$  &   $Q=0$  &  average \\ \hline 
$\sqrt{t_0}/a$(impr.) &  5.955(142) & 6.091(231) & 6.054(211) & 6.045(106)\\
$\sqrt{t_2}/a$(impr.) &  4.323(79) & 4.367(98) & 4.356(98) & 4.347(29)\\
$w_0/a$(impr.) & 6.998(325) & 7.490(438) & 7.304(561) & 7.297(333)\\
$w_2/a$(impr.) & 5.770(203) & 6.064(480) & 5.986(343) & 5.962(201)\\
\hline  \hline 
\end{tabular}
\caption{The gradient flow scales $\sqrt{t_i}/a$ and $w_i/a$ in lattice units determined from the improved discretization at different $\beta=$8.0 , 8.2, 8.4 in the different topological sectors.}
\label{tab:flowscales-stream}
\end{table*}

\begin{figure*}[tbh]
\centerline{
\includegraphics[width=0.5\textwidth]{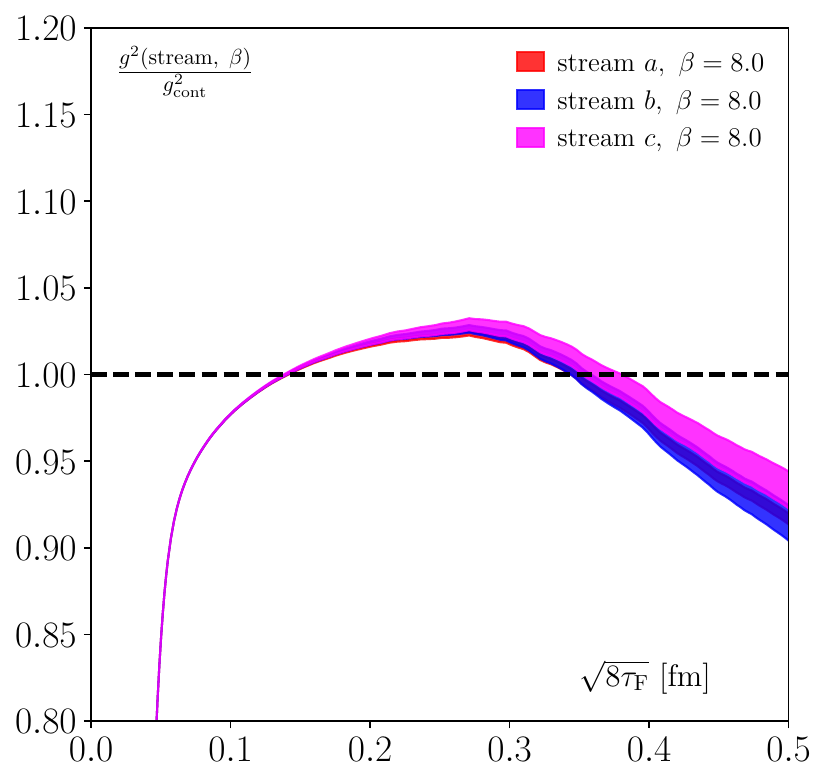}
\includegraphics[width=0.5\textwidth]{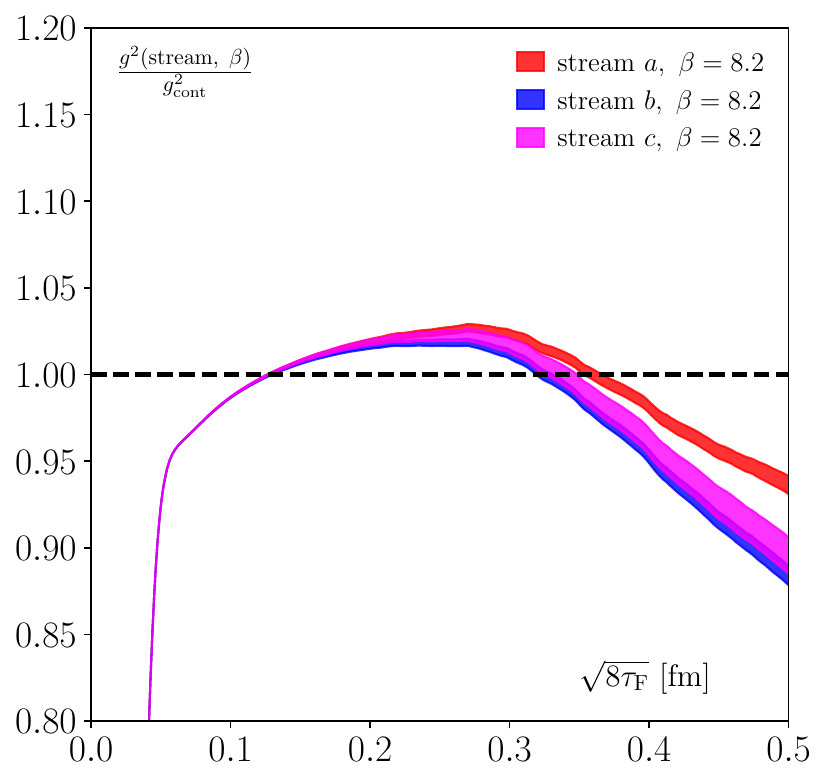}
}
\centerline{
\includegraphics[width=0.5\textwidth]{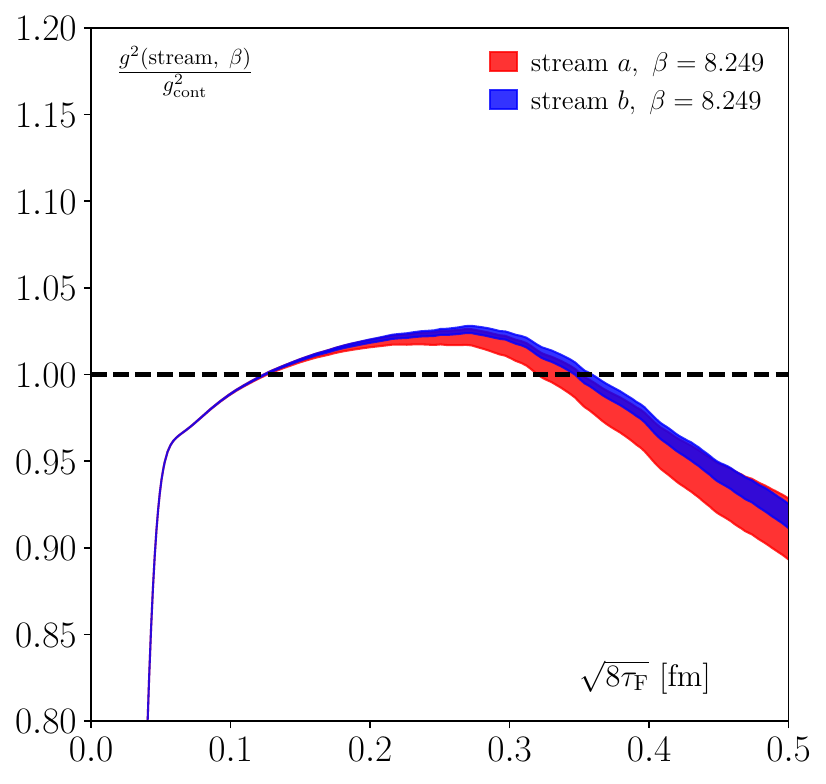}
\includegraphics[width=0.5\textwidth]{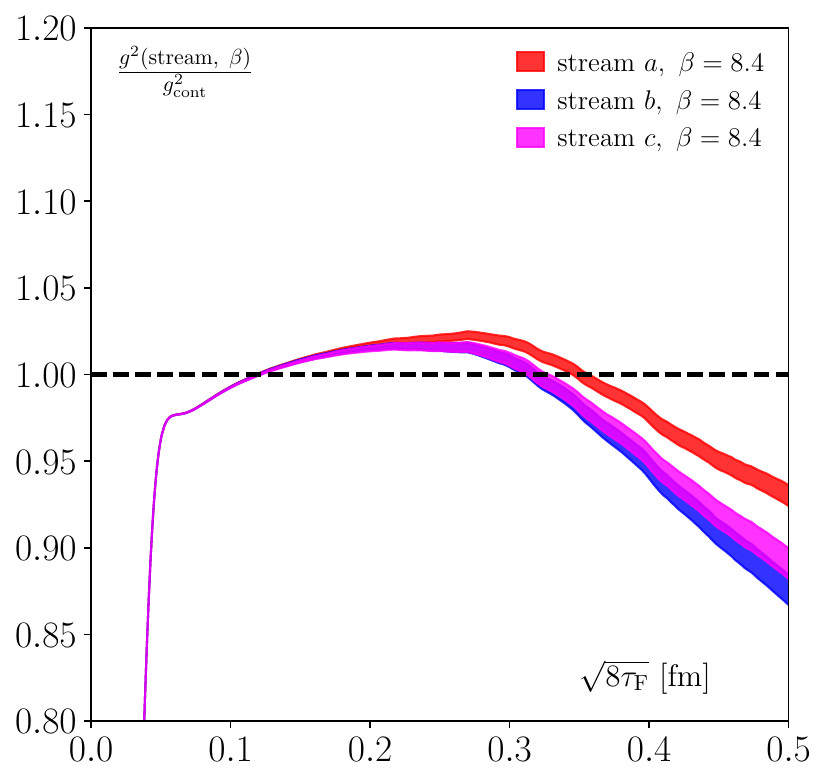}
}
\caption{Ratio of \(g_{\mathrm{flow}}^2\) obtained from individual streams to the continuum-extrapolated result, shown as a function of the flow time. In forming the ratio, the uncertainties of the continuum-extrapolated result are neglected. The continuum result is obtained from the \(m_s/m_l=5\) ensembles and corresponds to the brown curve in Fig.~\ref{fig:cont-extrap-summary}.}
\label{fig:g2-topo}
\end{figure*}

\begin{figure*}[tbh]
\centerline{
\includegraphics[width=0.5\textwidth]{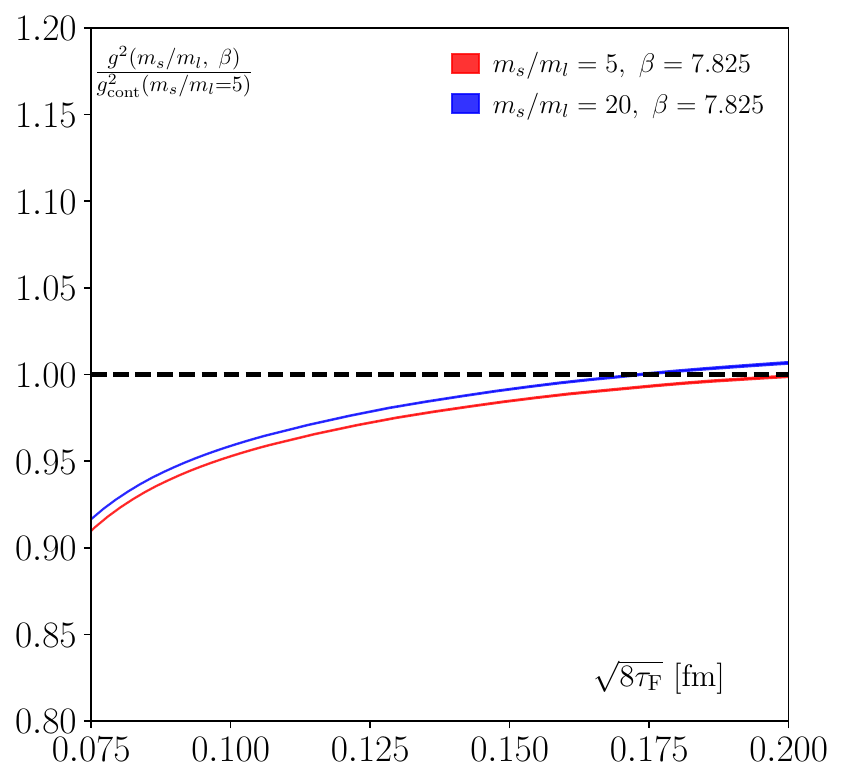}
}
\caption{Comparison of the ratios of \(g_{\mathrm{flow}}^2\) obtained from the \(m_s/m_l=5\) and \(m_s/m_l=20\) ensembles at \(\beta=7.825\) to the continuum-extrapolated result, shown as a function of the flow time. The data are computed using the improved discretization of the action. In forming the ratios, the uncertainties of the continuum-extrapolated result are neglected. The continuum result is obtained from the \(m_s/m_l=5\) ensembles and corresponds to the brown curve in Fig.~\ref{fig:cont-extrap-summary}.}
\label{fig:g2-mass}
\end{figure*}

In Figs.~\ref{fig:g2-topo} and \ref{fig:g2-mass}, we show the ratio of the gradient-flow coupling to the continuum-extrapolated result as a function of \(\sqrt{8\tauf}\). The denominator is the same in both figures, namely the continuum-extrapolated result obtained from the \(m_s/m_l=5\) ensembles, and its uncertainty is neglected when forming the ratio.
In Fig.~\ref{fig:g2-topo}, we show the results for different streams from the $m_s/m_s=5$ ensembles. While some differences in the value of the gradient-flow coupling are visible at large flow times, the results obtained from different streams agree within statistical uncertainties for \(\sqrt{8\tau_{\rm F}}<0.2~{\rm fm}\), which is the focus of the present study. We therefore combine the streams in our final analysis, assuming that the topology-related effect is smaller than the final quoted uncertainty.
In Fig.~\ref{fig:g2-mass}, we show the corresponding comparison at \(\beta=7.825\) for two values of the light-quark mass. As one can see from the figure, the results for \(m_s/m_l=5\) and \(m_s/m_l=20\) are not identical, but the difference is small in the flow-time range considered here. Thus, we can use the \(m_s/m_l=5\) lattice results to perform the comparison with perturbation theory.
From Figs.~\ref{fig:g2-topo} and \ref{fig:g2-mass}, we find that both the topology-related effects and the light-quark-mass effects are much smaller than the finite-lattice-spacing effects. This indicates that the dominant systematic effect in these comparisons is due to lattice discretization, rather than topological freezing or the light-quark-mass dependence.

\bibliographystyle{apsrev4-1}
\bibliography{references}

\end{document}